\documentclass[journal,twoside]{IEEEtran}
\renewcommand{\IEEEQED}{}

\newcommand{\norm}[1]{\left\lVert#1\right\rVert}

\ifCLASSINFOpdf
\else
\fi
\usepackage{xfrac}
\usepackage{amsmath}
\usepackage{multirow}
\usepackage{color}
\usepackage{amssymb}
\DeclareMathOperator*{\argmin}{argmin}

\usepackage{amsthm}
\usepackage{lipsum}
\usepackage{mathtools}
\usepackage{cuted}
\usepackage{float}

\usepackage{cite}
\begin{document}
\title{Performance Analysis of SSK-NOMA}


\author{Ferdi~Kara,~\IEEEmembership{Student Member,~IEEE,}
        Hakan~Kaya~%\IEEEmembership{Fellow,~OSA,}
        %and~Jane~Doe,~\IEEEmembership{Life~Fellow,~IEEE}% <-this % stops a space
\thanks{ This work is supported by the Scientific and Technological Research Council of Turkey (TUBITAK) under the 2211-E program}
\thanks{The authors are with the Department
of Electrical and Electronics Engineering, Zonguldak Bulent Ecevit University, Zonguldak, 67100 TURKEY e-mail: ( \{f.kara,hakan.kaya\}@beun.edu.tr).}% <-this % stops a space% <-this % stops a space
}

\maketitle

\begin{abstract}
In this paper, we consider the combination between two promising techniques: space-shift keying (SSK) and non-orthogonal multiple access (NOMA) for future radio access networks. We analyze the performance of SSK-NOMA networks and provide a comprehensive analytical framework of SSK-NOMA regarding bit error probability (BEP), ergodic capacity and outage probability. It is worth pointing out all analysis also stand for conventional SIMO-NOMA networks. We derive closed-form exact average BEP (ABEP) expressions when the number of users in a resource block is equal to i.e., $L=3$. Nevertheless, we analyze the ABEP of users when the number of users is more than i.e., $L\geq3$, and derive bit-error-rate (BER) union bound since the error propagation due to iterative successive interference canceler (SIC) makes the exact analysis intractable. Then, we analyze the achievable rate of users and derive exact ergodic capacity of the users so the ergodic sum rate of the system in closed-forms. Moreover, we provide the average  outage probability of the users exactly in the closed-form. All derived expressions are validated via Monte Carlo simulations and it is proved that SSK-NOMA outperforms conventional NOMA networks in terms of all performance metrics (i.e., BER, sum rate, outage). Finally, the effect of the power allocation (PA) on the performance of SSK-NOMA networks is investigated and the optimum PA is discussed under BER and outage constraints.  
\end{abstract}
\begin{IEEEkeywords}
NOMA, SSK, bit error rate, ergodic rate, outage performance
\end{IEEEkeywords}
% \fi
\IEEEpeerreviewmaketitle
\section{Introduction}
\IEEEPARstart{N}{on}-Orthogonal Multiple Access (NOMA) has emerged as a spectral efficient multiple access technique for future radio access networks and has received tremendous attention from the industry and academia. It is seen as one of the most strong candidates for 5G and beyond networks due to the ability of serving massive machine type communication (MMTC) \cite{Poor2015, Shin2017, Islam2017, Papers2017} and has already taken place in 3GPPP standards \cite{Specification2008, 3GPP2016}. 

NOMA is firstly proposed for next generation networks in \cite{Saito2013} and its superiority to orthogonal multiple access (OMA) counterparts  is proved in terms of overall system capacity. The outage performance of NOMA is analyzed when the users are randomly deployed to emphasize the effect of path loss \cite{Ding2014}. Then, the fairness between NOMA users is pointed out \cite{Timotheou2015} and optimum power allocation (PA) algorithms are investigated for NOMA when various constraints are considered \cite{Cui2016, Lei2016, Oviedo2017}. To show the effect of the inter-user-interference (IUI) on the error performance of users, an exact bit-error-rate (BER) analysis of NOMA is provided on two user single-input single-output (SISO) downlink and uplink networks \cite{Kara2018c}, \cite{Kara2018d}. In addition, BER union bound for multi-user downlink SISO networks is analyzed \cite{Bariah2018}. 

Due to its potential, the integration of NOMA with the other physical layer techniques (e.g. cooperative communication \cite{Kim2015a, Liu2015b, Kara2018a, Ding2016c, Liu2018, Kara2019}, multiple-input and multiple-output (MIMO) \cite{Sun2015}) has also received great attention from researchers since its implementation into well-known techniques is easily accomplished. The combination of cooperative communication and NOMA  is considered in three manners: NOMA using in cooperative communication \cite{Kim2015a}, cooperation within NOMA users \cite{Liu2015b, Kara2018a}, relay-aided cooperation for NOMA users \cite{Liu2018,Kara2019,Ding2016c}. Moreover, for the green communication concept, energy harvesting and wireless power transfer are considered for NOMA networks \cite{Liu2016c}. As more timely topics, cognitive radio in NOMA and NOMA in unnamed aerial vehicle (UAV) networks have attracted attention of researchers. In the above-mentioned studies, NOMA involved systems are analyzed mostly in terms of outage probability and the ergodic sum rate and the superiority of NOMA to related OMA networks is proved. 

\subsection{Related Works and Motivation}
Spatial modulation (SM) \cite{Mesleh2008} and space-shift keying (SSK) \cite{Jeganathan2009} are considered as two of other promising techniques for future wireless networks. Hence, the combination of SM/SSK with NOMA has also received attention. NOMA in MIMO-SM with finite alphabet inputs is proposed in \cite{Wang2017a} to enhance spectral efficiency of SM networks. Where users are grouped as only two users in a resource block and the symbols of users are transmitted by two transmit antennas simultaneously after SM is adopted for users' data streams. The mutual information is analyzed to obtain overall spectral efficiency and the superiority of the proposed NOMA-MIMO-SM to MIMO-OMA and MIMO-NOMA networks is presented. The proposed NOMA-MIMO-SM transmits signal with two antennas at the same time interval which needs two radio frequency (RF) chains so that the power consumption is increased. In addition, activating two transmit antennas at the same time interval increases the channel correlations. These situations are not considered in the paper. Then, in \cite{Chen2017b}, authors propose a combination of SM and NOMA in a cooperative network where infrastructure-to-vehicle (I2V), vehicle-to-vehicle (V2V) and intra-vehicle (vehicle to mobile user in car) are all included, over Rician fading channels. The network model is considered in two phases: In the first phase, NOMA is implemented for I2V and in the second phase, one of the vehicle acts as a relay for the user-in-car and for the other vehicle (V2V). In the second phase of communication, the relay implements NOMA after implementing SM for other vehicle's data stream.The mutual information is analyzed and the upper bound for spectral efficiency is derived. The authors also present BER performance of proposed model via simulations, however no analytical derivations are provided for BER. The proposed models in \cite{Wang2017a,Chen2017b} attract attention to implementation of SM and NOMA together to enhance spectral efficiency of well-known SM/NOMA networks. However, these models come with no solution to NOMA's drawbacks such as IUI.

Despite the superiority of NOMA involved networks to OMA networks in terms of sum rate and outage performance, NOMA networks are interference-limited. This IUI in NOMA causes a poor error performance for users compared to OMA networks \cite{Kara2018d}. Hence, the studies in the literature \cite{Poor2015, Islam2018} and the wireless standards \cite{Specification2008, 3GPP2016} mostly consider only two NOMA users in a resource block. The increase in users within a resource block causes dramatical growth of IUI and it limits the advantage of NOMA. In addition, it increases the number of required SIC processes in the receivers which cost high computational complexity. 
To overcome the decay in error performance of NOMA, SM aided multiple antenna network is proposed in \cite{Zhong2018}. The proposed model is based on applying SM principle for different users' data streams by selecting transmit antenna according to one of user's data and transmitting M-ary modulated symbol of the other user on selected antenna. The model is considered as NOMA since the BS communicates with two users at the same time and frequency block. Ergodic sum rate simulations are provided for the proposed system. The system model in \cite{Zhong2018} is later called as spatial multiple access (SMA) and the analytical performance analysis is provided \cite{Kara2018} to show its superiority to conventional two user NOMA networks. The SM aided NOMA network in \cite{Zhong2018} is then expanded for cooperative networks where one of the user acts as a relay for the other user and it is called as cooperative relaying system -SM-NOMA (CRS-SM-NOMA) \cite{Li2019}. Ergodic sum rate analysis is provided for CRS-SM-NOMA and it is shown that CRS-SM-NOMA is superior to both CRS-OMA and CRS-NOMA. 

Although the aforementioned studies \cite{Zhong2018, Kara2018, Li2019} achieve better performance than conventional NOMA systems, they still serve only two users in a resource block. To increase number of the served users in a orthogonal block (i.e., time, frequency, code), SSK combination with the NOMA is considered in \cite{Irfan2016} where the cell-edge user is assigned into spatial domain and two users are multiplexed with NOMA. Then, the authors in \cite{Kim2018} have extended \cite{Irfan2016} for more than three users where cell-edge users are allocated to spatial domain (i.e., antenna index) and the intra-cell users are served by conventional NOMA and it is called as Generalized SSK-NOMA (GSSSK-NOMA). In \cite{Kim2018}, the complexity analysis for GSSK-NOMA and BER simulations for cell-edge user is given. However, ML detections for cell-edge users provided in \cite{Irfan2016, Kim2018} have unrealistic assumptions such as superimposed NOMA signal is perfectly known at the cell-edge users. Hence, the provided detectors and the analysis turn out to be well-known conventional GSSK networks rather than GSSK-NOMA networks. Since the transmitted symbol in SSK-NOMA is complex, rather than only transmitting SNR such in conventional GSSK networks, a new detector for cell-edge user in SSK-NOMA should be provided. In addition, BER simulations are presented for only cell-edge user, neither BER simulations for NOMA users nor the outage/capacity analysis for any users are regarded. 

Furthermore, not only in SSK-NOMA but also in conventional NOMA networks, the error probability analyses are very limited in the literature. To the best of authors' knowledge, BER analysis of two user NOMA is given in \cite{Kara2018d} and the union BER bound is provided for multi-user NOMA in \cite{Bariah2018}. However, these works only consider SISO networks and there has been no work which consider multiple antenna models, yet. Since the intra-cell users are multiplexed by NOMA in the considered SSK-NOMA network, the analysis in this paper is the first study which considers multiple antenna NOMA networks in terms of error probability.
\begin{figure*}[!h]
    \centering
    \includegraphics[width=16cm,height=7cm]{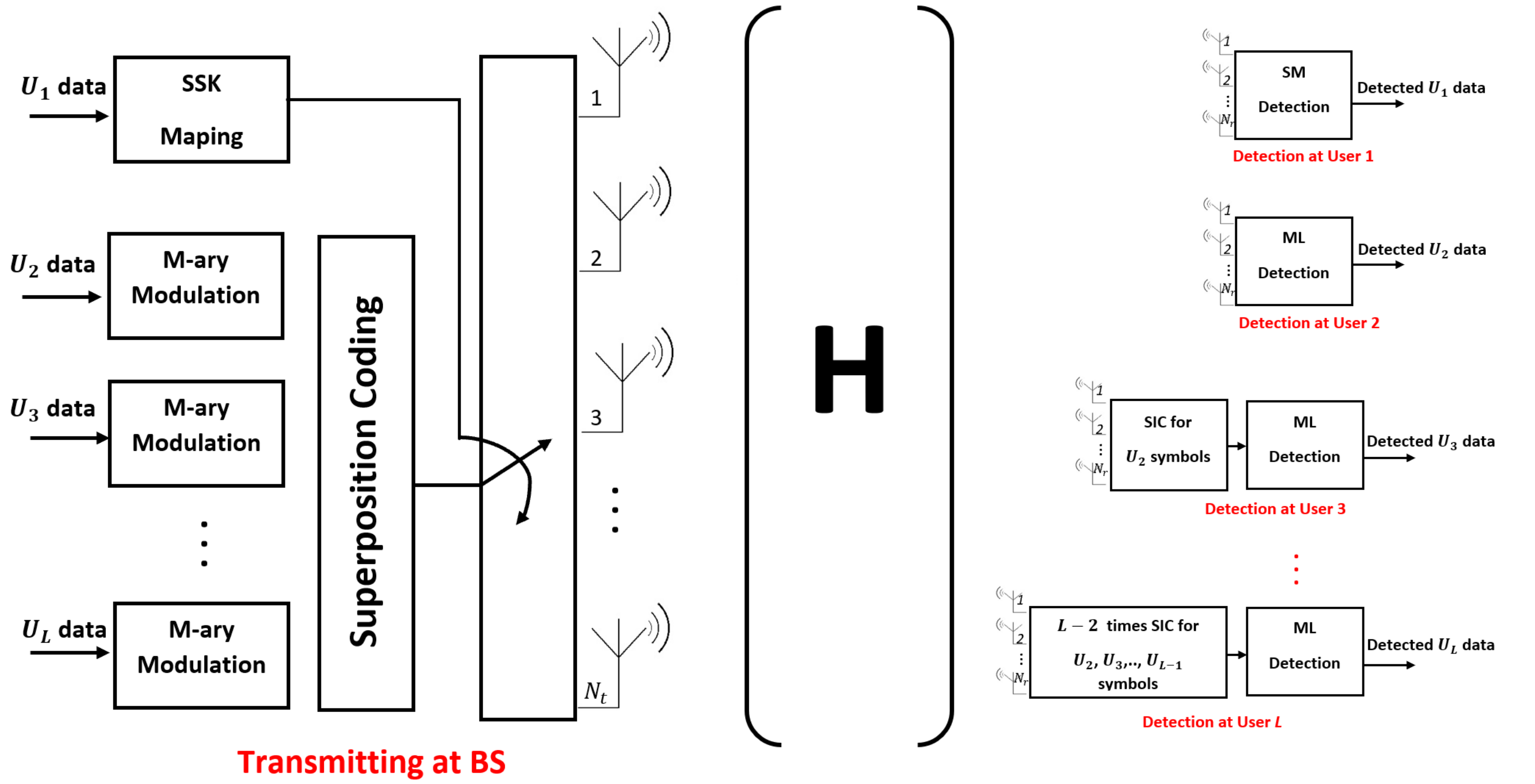}
    \caption{The illustration of SSK-NOMA}
    \label{fig1}
\end{figure*}
\subsection{Contribution}
The main contributions of this paper are as follows
\begin{itemize}
\item{SSK-NOMA aims to decrease IUI among NOMA users by assigning cell-edge user into spatial domain. Hence, all users (cell-edge and intra-cell) encounter much less IUI and have better performance (i.e., BER, sum rate, outage) than conventional NOMA networks. The IUI limits the number of users in a resource block (mostly by two users \cite{Specification2008, 3GPP2016}). SSK-NOMA provides usage of one more user in a resource block under the same performance requirements of NOMA.}  
\item{An optimum ML detector for cell-edge user in SSK-NOMA is presented and the complexity analysis for proposed detector is provided. The proposed detector does not require the knowledge of transmitted superimposed signal for NOMA users as referenced in \cite{Kim2018}, hence this paper represents more realistic scenarios for wireless communications. In addition, a tight BER bound for the given detector is derived in closed-form.}
\item{The closed-form exact BER analysis is provided for NOMA users when the total number of users is equal to $L=3$. In addition, a BER union bound is derived when the number of users is more than $L=3$ and arbitrary modulation constellation is considered. These analyses also stand for conventional SIMO-NOMA networks and this is also the first study which investigates the error performance of multiple antenna NOMA networks to the best of authors' knowledge.}
\item{The ergodic capacity of the users and the ergodic sum rate of the overall system are analyzed and exact expressions are provided in closed-form. The average outage probability of the SSK-NOMA is also derived in the closed-form. These analyses provide insights for analysis of uplink conventional NOMA networks.}
\item{We present extensive simulations to validate all derived theoretical analysis. Based on the simulation results, we reveal that SSK-NOMA is superior to conventional NOMA in terms of all performance metrics and it allows to serve more than three users in a resource block without decreasing the performance of network which is very promising for massive type communication. In addition, limiting the required RF chain with one reduces the energy consumption at the transmitter which is very important for the energy efficiency so that for the green communication concept of 5G}
\end{itemize}
\subsection{Organization}
The remainder of this paper is as follows. In Section II, the system and the channel models are introduced and the optimum detectors for the users are provided. The complexity analysis for the receivers in SSK-NOMA and comparison with the conventional NOMA networks are provided. In Section III, the performance analysis of SSK-NOMA is derived. Then, the evaluation of the derivations in Section III are presented via Monte Carlo simulations in Section IV. The discussion for the effect of the PA on the performance of SSK-NOMA is also given in this section. Finally, the results are discussed and the paper is concluded in Section V.
\subsection{Notation}
The bold uppercase and lowercase letters denote the matrices and the vectors, respectively. We use $(.)^T$ for transpose, $(.)^H$ for conjugate transpose and $\left||.|\right|_F$ for the Frobenius form of a matrix/vector. We use $\left|.\right|$ for the absolute value of a scalar/vector and $\binom{.}{.}$ for the binomial coefficient. $\hat{\ }$ denotes the estimated symbol or index. $Re\{\}$ is the real component of a complex symbol or vector. $P_r(A)$ denotes the probability of the event $A$ whereas $P_r(A|B)$ is the probability of the event $A$ under the condition that $B$ has already occurred. 
\section{System and Channel Models}
We consider the usage of SSK-NOMA with one base station (BS) and $L$ users (i.e., $U_1,\ U_2,\ U_3,\ \dots\ U_L$) within a resource block\footnote{$L$ users are assumed to be assigned into a resource block. In each resource block, users with different numbers can be served and each resource block is on an orthogonal domain (frequency, time, code), hence the resource allocation algorithms for NOMA can be easily adopted in SSK-NOMA} in a MIMO downlink network. The BS and users are equipped with $N_t$ and $N_r$ antennas, respectively. The channel between BS and each user is represented as $\mathbf{H_i} \subset \mathbb{C}^{N_t\times N_r}$. $\mathbf{H_i}$ consists of $\mathbf{h_{i,j}}=\left[h_{i,j,1},\ {h_{i,j,2}}, \dots,h_{i,j,N_r}\right]^T, \quad j=1,2,\dots,N_t$ vectors. For user $i$,  the channel gains between each transmit antenna and each receive antenna ($ h_{i,j,k}$) are assumed to be independent and identically distributed (i.i.d) as $CN(0,\sigma_i^2)$. The channels are modeled as combination of large-scale fading and small-scale fading (flat fading). $\sigma_i^2$ denotes the large-scale fading coefficient determined by the distance of the $i$th user to the BS. Without loss of generality, we assume the users are sorted in ascending order according to their distance to the BS so that the average channel gains i.e. $\sigma_1^2\leq\sigma_2^2\leq\sigma_3^2\leq\dots\leq\sigma_L^2$. 

As shown in Fig. 1, SSK-NOMA integrates the SSK and NOMA techniques. In SSK-NOMA, one of the users is assigned into spatial domain and determines transmit antenna index to be activated whereas the other users are multiplexed by NOMA. The users ($i\geq 2$) are multiplexed in power domain with the different power allocation (PA) (i.e., $a_i$) coefficients. The superposition coded symbol of the users ($i\geq2$) is given by
\begin {equation}
\chi=\sum_{i=2}^{L}{\sqrt{a_i}s_i}
\end {equation}
where $s_i$ is the base-band complex symbol of the $i$th user and $a_i$ is PA coefficient for  the $i$ th user. $\sum_{i=2}^{L}{a_i}=1$ and $a_2>a_3>\dots>a_L$. Hence, the $\chi\subset\mathbb{C}^{M_ {T}}$ is defined, $M_{T}$ is the total signal dimension of SC symbols and it is given as $M_{T}=\prod_{i=2}^{L}{M_i}$, where $M_i$ is the modulation level of the $i$th user. The binary symbol of the $U_1$ determines the transmit antenna index (i.e., $j=v$) -which transmit antenna will be activated at BS-. The transmitted vector at BS contains only one non-zero element which is equal to $\chi$. The transmitted vector is
\begin{equation}
\begin{split}
&\mathbf{x}=[\overbrace{0 \ 0 \ 0 \ \dots \ \chi \dots \ 0 \ 0 \ 0 }^{N_t}]^T \\
& \ \ \ \ \ \ \ \ \  \  \ \  \  \ \ \ \ \ \uparrow v^{th} position
\end {split}
\end {equation}
and the received vector at each user is given
\begin {equation}
\mathbf{r_i}=\sqrt{P}\mathbf{H_i}\mathbf{x}+\mathbf{w_i}\ \ \ i=1,2,...L
\end{equation}
where $P$ is the transmit power and $\mathbf{w_i}$ is $N_r$-dim additive white Gaussian noise (AWGN) at user $i$ and each dimension is distributed as $CN(0,N_0)$. The users with $N_r$ receive antennas firstly implement a maximal ratio combining (MRC) and detect their own symbols. Since the cell-edge user and intra-cell users are assigned into spatial and power domain, respectively, the required detectors are implemented in different ways for those users.  
\subsection{Detection at the first user (cell-edge user-SM)}
The detection at the first user should be implemented to obtain transmit antenna index. Although the symbols of the first user are mapped into only transmit antenna index, a SM detector must be used at the first user rather than SSK detector since the non-zero element is equal to complex SC symbol as in SM rather than only the signal-to-noise ratio (SNR) value as in SSK. Nevertheless, the optimal SM detector \cite{Transmission2008} should be adopted for SSK-NOMA since the transmitted symbol by active antenna is the total SC symbol of NOMA users. Considering this, the optimal detector of cell-edge user is given as
\begin{equation}
\left[\hat{v}, \hat{\chi}_k\right]=\argmin_{v,k}{\sqrt{\rho}\norm{\mathbf{g_{v,k}}}_F^2-2Re\{\mathbf{{r_1}}^H\mathbf{g_{v,k}\}}},  
\end{equation}
where $v=1,2,..N_t$ and $k=1,2,\dots,\prod_{i=2}^{L}{M_i} $. $\mathbf{g_{v,k}={h_{1,v}}}\chi_k$ is defined and $\rho=\sfrac{P}{N_0}$ is the average SNR for each antenna. 
\subsection{Detection at the users for $i\geq2$ (intra-cell users-NOMA)}
The detections at the other users ($i\geq2$) should be implemented to obtain transmitted complex symbol of the related user. Since NOMA is implemented for the users i.e., $i\geq2$, iterative SIC implementations should be accomplished according to decoding order. Given that the second user has the highest allocated power coefficient, the detection for $U_2$ is obtained by treating as noise to the other users' symbols so that SIC is not required. Hence, the ML detection for the $U_2$ is given as
\begin{equation}
\hat{s}_2=\argmin_{n}{\norm{\mathbf{{r_2}}-\sqrt{a_2P}\mathbf{{h_{2,v}}}s_{2,n}}^2}, \quad n=1,2,..M_2,
\end{equation}
where $s_{i,n}$ denotes the $n$ th symbol within constellation $M_i$ of the $i$ th user. The users ($i\geq3$) should implement iterative SIC processes and eliminate the symbols of the users which are before in the detecting order ($m<i$). The ML detection is given as
\begin{equation}
\hat{s}_i=\argmin_{n}{\norm{\mathbf{{r'_i}}-\sqrt{a_iP}\mathbf{{h_{i,v}}}s_{i,n}}^2}, \quad n=1,2,..M_i,
\end{equation}
where $\mathbf{{r'_i}}$ is
\begin{equation}
\mathbf{{r'_i}}=\mathbf{{r_i}}-\sum_{m=2}^{i-1}\mathbf{{h_{i,v}}}\sqrt{a_mP}\hat{s}_m
\end{equation}
\subsection{Complexity}
To compare the computational receiver complexity of SSK-NOMA with conventional NOMA, we provide complexity analysis for the detections given in Section II-A and II-B. In the complexity analysis, we use the number of complex operations as the complexity metric. In SSK-NOMA, the cell-edge user (i.e, $U_1$) should implement SM-ML detection given in (4). With the aid of \cite{Transmission2008}, the complexity of the SM-ML detection (4) is expressed as 
\begin{equation}
\delta_{SM-ML}=2N_rN_t+N_tM_T+M_T
\end{equation} 
The receiver complexity for intra-cell users (i.e, $U_2, U_3, \dots, U_L$) depends on the ML detection (5), (6) and the number of SIC process (7) which should be implemented. Hence, we firstly derive number of required ML and SIC processes for users. Since $U_2$ has the highest PA coefficient, $U_2$ implements only ML detection (5) by considering the other users' signal as noise and no SIC process is required. On the other hand, the users (i.e., $i\geq 3$) should implement $i-2$ times iterative SIC processes (7) to subtract the other users' interference and then implements ML detection (6) to obtain own signal. Hence, the number total of ML detection to detect M-ary modulated signals is given as
\begin{equation}
\mathcal{O}_{M_{ary}-ML}^{SSK-NOMA}=L-1
\end{equation} 
and the number of required SIC processes is given as
\begin{equation}
\mathcal{O}_{SIC}^{SSK-NOMA}=\sum_{i=3}^L(i-2)=\frac{(L-2)(L-1)}{2}
\end{equation} 
With the aid of \cite{Kim2010}, the complexity of ML detector (5), (6) is expressed as
\begin{equation}
\delta_{M_{ary}-ML,i}=4N_rM_i \quad i=2,3,\dots, L
\end{equation} 
We note that the SIC processes at $i$th user (7) includes ML detections for users $m<i$. Hence, considering the ML complexity (11), the complexity of SIC processes at user $i$ is given as
\begin{equation}
\delta_{SIC,i}=\sum_{m=2}^{i-1}\left(4N_rM_m+2N_r\right) \quad m<i, \ i=3, 4, \dots, L
\end{equation} 
Considering the complexity of detections (8), (11), (12) and the number of the required operations (9), (10), the total receiver complexity for SSK-NOMA is derived as
\begin{equation}
\begin{split}
\delta_{SSK-NOMA}=&\underbrace{2N_rN_t+N_tM_T+M_T}_{SM\ detection} + \underbrace{\sum_{i=2}^{L}4N_rM_i}_{ML\ detection}\\
&+\underbrace{\sum_{i=3}^{L}\sum_{m=2}^{i-1}\left(4N_rM_m+2N_r\right)}_{SIC\ processes} 
\end{split}
\end{equation}
To provide a fair comparison, we analyze the complexity of conventional NOMA networks when the transmit antenna is one since the SSK-NOMA limits the required RF chain to only one by activating one transmit antenna according to cell-edge user's data. Otherwise, the complexity of NOMA would increase exponentially by the number of transmit antenna. To obtain total complexity of NOMA, we provide the numbers of required ML and SIC processes. Since all users are multiplexed by different PA coefficients, the numbers of ML and SIC processes in NOMA are given
\begin{equation}
\mathcal{O}_{M_{ary}-ML}^{NOMA}=L
\end{equation} 
and
\begin{equation}
\mathcal{O}_{SIC}^{NOMA}=\sum_{i=2}^L(i-1)=\frac{L(L-1)}{2}
\end{equation}
Recalling the complexity of ML (11) and SIC processes (12), the total receiver complexity for NOMA networks is derived as
\begin{equation}
\delta_{NOMA}=\underbrace{\sum_{i=1}^{L}4N_rM_i}_{ML\ detection} +\underbrace{\sum_{i=2}^{L}\sum_{m=1}^{i-1}\left(4N_rM_m+2N_r\right) }_{SIC\ processes} 
\end{equation}
We present a complexity comparison of SSK-NOMA and conventional NOMA for different scenarios in Table I. We assume modulation levels are same for all users in NOMA ($M_i=M$) and for fairness in terms of first user's data rate, the number of transmit antenna in SSK-NOMA is also equal to modulation level ($N_t=M$). One can see that, SSK-NOMA has less complexity than NOMA when the constellation size of users ($M_i$) is relatively low whereas for higher modulation levels, SSK-NOMA requires more complex operations since the total size of SC symbols ($M_t$) becomes larger and it increases receiver complexity at the cell-edge user. However, considering the performance gain of SSK-NOMA presented in Section IV, this complexity is considered as affordable. In addition, the other users (intra-cell) encounter much less receiver complexity since the number of required SIC processes becomes less. 
\begin{table}[!h]
  \centering
  \caption{\textsc{Complexity comparison of SSK-NOMA with NOMA}}
  \label{table1}
  \begin{tabular}{|c|c|c|c|c|}
	\hline
	\multirow{2}{*}{$L$}&\multirow{2}{*}{$M$}&\multirow{2}{*}{$N_r$}&\multicolumn{2}{c|}{Complexity} \\ \cline{4-5}
	& & & $\delta_{SSK-NOMA}$&$\delta_{NOMA}$ \\
	\hline
	3&2&2&72&108 \\
	\hline
	3&4&4&312&408 \\
	\hline
	4&2&2&140&184 \\
	\hline
	4&4&4&760&688\\
	\hline
	5&2&2&240&280 \\
	\hline
	5&4&4&2000&1040 \\
	\hline
	\end{tabular}
\end{table}
\section{Performance Analysis} 
In this section, we provide analytical performance analyses of SSK-NOMA over Rayleigh fading channels in terms of bit error probability (BEP), ergodic sum rate and outage probability. All analyses are presented in two parts as for cell-edge user and for intra-cell users since they are assigned into different domains (i.e., spatial and power) and the detections at those users (4), (5), (6) are implemented differently.
\subsection{Average Bit Error Probability (ABEP) Analysis} 
\subsubsection{ABEP for the first user ($U_1$)}
The exact bit-error probability for the SM systems cannot be derived. Nevertheless, a tight upper bound can be obtained by using well-known union bounding technique. Hence, the error performance of the optimal detector of $U_1$ in SSK-NOMA given in (4) is determined by utilizing \cite[Eq. (5)]{Transmission2008}
\begin{equation}
\ P(e)\le\sum_{v=1}^{N_t}\sum_{\hat{v}=1}^{N_t}\sum_{k=1}^{M_{T}}\sum_{\hat{k}=1}^{M_{T}}\frac{N(\chi_{k}\rightarrow \hat{\chi}_{k})PEP(\mathbf{x}_{v,k}\rightarrow \mathbf{x}_{\hat{v},\hat{k}})}{M_{T}N_t},
\end{equation}
where $\mathbf{x}_{v,k}$ defines the vector $\mathbf{x}$ which has the SC symbol of $\chi_k$ at the $v$ th index. $N(\chi_{k}\rightarrow \hat{\chi}_{k})$ is the Hamming distance between $\chi_k$ and $\hat{\chi}_{k}$. $PEP(\mathbf{x}_{v,k}\rightarrow \mathbf{x}_{\hat{v},\hat{k}})$ is defined as the pairwise error probability (PEP) when the $\mathbf{x}_{v,k}$ is transmitted and it is estimated as $\mathbf{x}_{\hat{v},\hat{k}}$. Recalling that the $U_1$ takes the only estimated transmit antenna index as the output of SM detection given in (4), the union bound expression (17) is simplified as
\begin{equation}
\ P(e)\le\sum_{v=1}^{N_t}\sum_{\hat{v}=1}^{N_t}\frac{PEP(\mathbf{x}_{v,k}\rightarrow \mathbf{x}_{\hat{v},\hat{k}})}{N_t},
\end{equation}
To derive PEP expression in (18), the conditioned PEP on $\mathbf{H_1}$ is given in \cite[Eq. (6)]{Transmission2008} as $Q(\sqrt{\kappa})$\footnote{However, the PEP expression is only given for real constellations (i.e., BPSK)} where $\kappa\triangleq\frac{\rho}{2}||\mathbf{g_{v,k}}-\mathbf{g_{\hat{v},\hat{k}}}||_F^2$. For general constellations, the PEP given \cite[Eq. (7)]{Transmission2008} is adopted by utilizing \cite[Eq. (7)]{Renzo2012}, \cite[Eq. (64)]{Alouini1999} and the PEP is derived for the Rayleigh fading channels as 
\begin{equation}
\begin{split}
\ &PEP(x_{v,k}\rightarrow x_{\hat{v},\hat{k}})=\\
&{\mu_1}^{N_r}\log_2{M_T} \sum_{\lambda=0}^{N_r-1}\binom{N_r-1+\lambda}{\lambda}\left(1-\mu_1\right)^\lambda,
\end{split}
\end{equation}
where $\mu_{1}=\frac{1}{2}\left(1-\sqrt{\frac{\sigma_a^2}{{2+\sigma}_a^2}}\right)$ and $\sigma_a^2=\frac{\rho{\sigma_1}^2\left(\left|\chi_{k}\right|^2+\left|\hat{\chi}_{k}\right|^2\right)}{4}$. It is worth noting that the $\sigma_a^2$ depends on the transmitted $\chi_k$ and estimated $\hat{\chi}_{k}$ symbols. Since the SC is applied for the $\chi_k$ symbols, the energy level of the symbols is not constant and changes according to the PA coefficients. Hence, the PEP given in (19) should be averaged considering all scenarios. Substituting (19) into (18), the ABEP of the $U_1$ is derived 
\begin{equation}
\ P_1(e)\le { \frac{N_t}{2} {\mu_1}^{N_r}\log_2{M_T}}\sum_{\lambda=0}^{N_r-1}\binom{N_r-1+\lambda}{\lambda}\left(1-\mu_1\right)^\lambda
\end{equation}
\subsubsection{For users $i\geq2$}
The ABEP expression is highly dependent to the modulation constellation chosen. In addition, since SC is applied at the BS for symbols of users for $i\geq2$, iterative SIC process is required at the users for $i\geq3$. In the presence of imperfect SIC, an error propagation from SIC process to symbols of users with higher detection order occurs and the exact error analysis becomes intractable when $L>3$. Hence, we analyze exact ABEP for $L=3$ and QPSK is used for both users. Then an upper bound for $L>3$ and arbitrary constellation is derived.
\paragraph{Exact-ABEP analysis for $L=3$}\newtheorem{proposition}{Proposition}
\begin{proposition}Considered the QPSK is chosen for both NOMA users (i.e., $U_2$ and $U_3$) and Gray mapping is applied, the conditional BEP of $U_2$ is
\begin{equation}
\ P_2(e|\gamma_2)=\sum_{c=1}^{2}\frac{1}{2}Q\left( \sqrt{\zeta_c\gamma_2}\right)
\end{equation}
\end{proposition}
\begin{IEEEproof}
See Appendix A
\end{IEEEproof}
where $\zeta_c,\ c=1,2$ denotes the base-band energy levels of the SC $\chi_k$ symbols. $\zeta_1=\left(\sqrt{a_2}-\sqrt{a_3}\right)^2$ and  $\zeta_2=\left(\sqrt{a_2}+\sqrt{a_3}\right)^2$ are defined.  $\gamma_2$ is the SNR at the output of MRC and in case Rayleigh fading, it follows Chi-square distribution with $2N_r$ degree of freedom and the probability density function (PDF) is given as \cite[Eq. (2.3-21)]{Proakis2008} 
\begin{equation}
\ p_{\gamma_{2}}(\gamma_{2})=\frac{{\gamma_{2}}^{N_r-1}e^{\sfrac{-{\gamma_{2}}}{\overline{\gamma}_{2}}}}{\Gamma(N_r){\overline{\gamma}_{2}}^{N_r}}
\end{equation}
where $\overline{\gamma}_{2}=\rho\sigma_2^2$. The ABEP of $U_2$ is obtained by $\int_{0}^{\infty}P_2(e|\gamma_2)p_{\gamma_2}(\gamma_2)d\gamma_2$. With the aid of \cite[Eq. (9.6)]{Simon2004}, the ABEP of $U_2$ is derived as
\begin{equation}
\ P_2(e)=\sum_{c=1}^{2}\frac{\left(\frac{1-\mu_2}{2}\right)^{N_r}}{2} \sum_{\lambda=0}^{N_r-1}\binom{N_r-1+\lambda}{\lambda}\left(\frac{1+\mu_2}{2}\right)^\lambda,
\end{equation}
where $\mu_2=\sqrt{\frac{\zeta_c\overline{\gamma}_{2}}{2+\zeta_c\overline{\gamma}_{2}}}$

In order to obtain BEP for the $U_3$ we should consider two cases: whether the symbols of $U_2$ are detected correctly or erroneously during the SIC process at $U_3$ \cite[Eq. (13)]{Kara2018d}. Hence, the BEP of $U_3$ is given as
\begin{equation}
\ P_3(e)=P_3(e|correct_{U_2})+P_3(e|error_{U_2})
\end{equation}
\begin{proposition}In the first case, we assume that symbols of $U_2$ are detected correctly and subtracted from the total received signal at $U_3$. In this case, the BEP of $U_3$ is given as
\begin{equation}
\ P_3(e|correct_{U_2})=\frac{1}{2}\left[2Q\left( \sqrt{\zeta_3\gamma_3}\right)-Q\left( \sqrt{\zeta_2\gamma_3}\right)\right]
\end{equation}
\end{proposition}
\begin{IEEEproof}
\ {See Appendix B}
\end{IEEEproof}
where $\zeta_3\triangleq a_3$
\begin{proposition}In the second case, we assume that the symbols of $U_2$ are detected erroneously and subtracted from the received signal at $U_3$. In this case, the BEP of $U_3$ is given as
\begin{equation}
\begin{split}
\ &P_3(e|error_{U_2})= \\
&\frac{1}{2}\left[Q\left( \sqrt{\zeta_1\gamma_3}\right)-Q\left( \sqrt{\zeta_4\gamma_3}\right)+Q\left( \sqrt{\zeta_5\gamma_3}\right)\right]
\end{split}
\end{equation}
\end{proposition}
\begin{IEEEproof}
See Appendix C
\end{IEEEproof}
where $\zeta_4=\left(2\sqrt{a_2}-\sqrt{a_3}\right)^2$ and $\zeta_5=\left(2\sqrt{a_2}+\sqrt{a_3}\right)^2$
Substituting (25) and (26) into (24), we obtain the exact BEP of the $U_3$ is derived as in (27) (see the top of the next page).
\begin{figure*}[!t]
 \begin{equation}
\ P_3(e|\gamma_3)=\frac{1}{2}\left[2Q\left(\sqrt{\zeta_3\gamma_3}\right)+Q\left(\sqrt{\zeta_1\gamma_3}\right)-Q\left(\sqrt{\zeta_2\gamma_3}\right)-Q\left(\sqrt{\zeta_4\gamma_3}\right)+Q\left(\sqrt{\zeta_5\gamma_3}\right)\right]
\end{equation}
\hrulefill
\end{figure*}
By averaging  (27) over instantaneous $\gamma_3$, with the aid of (22) and \cite[Eq. (9.6)]{Simon2004} we obtain that ABEP of $U_3$ turns out to be
\begin{equation}
\begin{split}
\ &P_3(e)= \\
&\sum_{c=1}^{5}\frac{A_c\left(-1\right)^{c+1}\left(\frac{1-\mu_3}{2}\right)^{N_r}}{2} \sum_{\lambda=0}^{N_r-1}\binom{N_r-1+\lambda}{\lambda}\left(\frac{1+\mu_3}{2}\right)^\lambda
\end{split}
\end{equation}
where $A_c$ is a constant and is defined as $A_c=2 \ if,c=3, otherwise \ 1$ and $\mu_3=\sqrt{\frac{\zeta_c\overline{\gamma}_3}{2+\zeta_c\overline{\gamma}_3}}$.
% or
\paragraph{Union bound analysis for $L\geq3$}
The error propagation from iterative SIC implementations is intractable for $L\geq3$. Hence, we analyze union bound for BER of the users. To derive the union bound, PEP for each symbol is obtained and then averaged for all possible symbols. By utilizing PEP of SISO NOMA network \cite{Bariah2018}, conditional PEP for symbols of user $i$ (i.e, $i\geq2$) is given 
\begin{equation}
PEP(s_i\rightarrow\hat{s}_i|\mathbf{h_{i,v}})=Q\left(\frac{\beta_i\sqrt{\mathbf{h_{i,v}}\mathbf{h_{i,v}^H}}}{\vartheta}\right)
\end{equation}
where   
\begin{equation}
\begin{split}
\beta_i=&\sqrt{a_i\rho}\left|\triangle_i\right|^2+2[Re\underbrace{\{ \triangle_i\sum_{p=i+1}^L\sqrt{a_p\rho}s_p^*\}}_{noise \ term\ for \ p>i} \\
&+Re\underbrace{\{ \triangle_i\sum_{q=2}^{i-1}\sqrt{a_q\rho}\triangle_q^*\}}_{SIC \ errors\ for \ q<i}]
\end{split}
\end{equation}
where $\vartheta=\sqrt{2}\left|\triangle_i\right|$ and $\triangle_i=s_i-\hat{s}_i$ are defined (for proof see the \cite[Eq.(17)-Eq.(20)]{Bariah2018}). It is clearly seen that the SIC errors term is equal to zero for $i=2$ and the noise term is equal to zero for $i=L$. When we averaged the conditional PEP over instantaneous SNR by using PDF (22), we derive average PEP as
\begin{equation}
\begin{split}
\ &PEP(s_i\rightarrow\hat{s}_i)=\\
&\left(\frac{1-\xi_i}{2}\right)^{N_r}\sum_{\lambda=0}^{N_r-1}\binom{N_r-1+\lambda}{\lambda}\left(\frac{1+\xi_i}{2}\right)^\lambda,
\end{split}
\end{equation}
where $\xi_i=\sqrt{\frac{\sigma_i^2\beta_i^2}{2\vartheta^2+\sigma_i^2\beta_i^2}}$. It is worth noting that the PEP depends on the transmitted and detected symbols of the users, hence it should be averaged all possible symbols of users. BER union bound in terms of PEP is given
\begin{equation}
\begin{split}
\ BER_i^{union}\leq\sum_{s_i}N(s_i\rightarrow\hat{s}_i)\sum_{s_i\neq\hat{s}_i}PEP(s_i\rightarrow\hat{s}_i|s_p,\triangle_p)& \\
 \forall p\neq i&
\end{split}
\end{equation}
\subsection{Ergodic Sum Rate Analysis}
In this subsection, we analyze ergodic rate of each user by averaging achievable (Shannon) rate at each user over channel fading. Then, we obtain total ergodic capacity/sum rate of SSK-NOMA to show its superiority to conventional NOMA networks.
\subsubsection{Ergodic capacity of first user ($U_1$)}
Achievable rate of $U_1$ only depends on the number of the transmit antennas, since the binary symbols of the $U_1$ are mapped into transmit antenna index. Hence, achievable rate of $U_1$ is given as
\begin{equation}
R_1=\log_{2}{N_t}
\end{equation}
Nevertheless, the achievable rate of SSK is mostly assumed as the bits detected correctly in the literature. Hence, the achievable rate of $U_1$ can be given as
\begin{equation}
R_1=\log_{2}{N_t}\left[1-P_1(e|\mathbf{h_{1,v}})\right]
\end{equation}
where $P_1(e|\mathbf{h_{1,v}})$ is the BEP of $U_1$. To obtain the ergodic capacity of $U_1$, once we calculate the $\int_{0}^{\infty}R_1 p_{\gamma_1}(\gamma_1)d\gamma_1$, we obtain
\begin{equation}
C_1=\log_{2}{N_t}\left[1-P_1(e)\right]
\end{equation}
where $P_1(e)$ is the ABEP of $U_1$ and given in (20).
\subsubsection{Ergodic capacity for users $i\geq2$}
Achievable rate of the $i$ th (i.e., $i\geq2$) user is given by
\begin{equation}
R_i=\log_{2}{(1+SINR_i)}
\end{equation}
and the $SINR_i$ is defined as
\begin{equation}
SINR_i=\frac{a_i\rho\mathbf{h_{i,v}}\mathbf{h_{i,v}^H}}{1+\sum_{p=i+1}^{L}a_p\rho\mathbf{h_{i,v}}\mathbf{h_{i,v}^H}}
\end{equation}
substituting (37) into (36)
\begin{equation}
\begin{split}
R_i=&\log_{2}{\left(1+\frac{a_i\rho\mathbf{h_{i,v}}\mathbf{h_{i,v}^H}}{1+\sum_{p=i+1}^{L}a_p\rho\mathbf{h_{i,v}}\mathbf{h_{i,v}^H}}\right)} \\
=&\log_{2}{\left(\frac{1+\sum_{p=i}^{L}a_p\rho\mathbf{h_{i,v}}\mathbf{h_{i,v}^H}}{1+\sum_{p=i+1}^{L}a_p\rho\mathbf{h_{i,v}}\mathbf{h_{i,v}^H}}\right)}  \\
=&\log_{2}{\left(1+\sum_{p=i}^{L}a_p\rho\mathbf{h_{i,v}}\mathbf{h_{i,v}^H}\right)} \\
&-\log_{2}{\left(1+\sum_{p=i+1}^{L}a_p\rho\mathbf{h_{i,v}}\mathbf{h_{i,v}^H}\right)} 
\end{split}
\end{equation}
Recalling that $\rho\mathbf{h_{i,v}}\mathbf{h_{i,v}^H}$ is the SNR at the MRC output of $i$ th user, we obtain ergodic rate of the $i$ th user
\begin{equation}
\begin{split}
C_i=&\int_{0}^{\infty}\log_{2}{\left(1+\sum_{p=i}^{L}a_p\rho\gamma_i\right)}p_{\gamma_i}(\gamma_i)d\gamma_i \\
&-\int_{0}^{\infty}\log_{2}{\left(1+\sum_{p=i+1}^{L}a_p\rho\gamma_i\right)} p_{\gamma_i}(\gamma_i)d\gamma_i
\end{split}
\end{equation}
\begin{figure*}[!t]
\begin{equation}
\begin{split}
C_i=\sum_{t=1}^{2}(-1)^{t-1}\frac{\log_2{e}}{\Gamma(N_r)}\sum_{\lambda=0}^{N_r-1}\frac{(N_r-1)!}{(N_r-1-\lambda)!}
\left[\frac{(-1)^{N_r-\lambda-2}}{{\eta_t}^{N_r-1-\lambda}}e^{\sfrac{1}{{\eta_t}}}\mathbf{Ei}\left(-\frac{1}{{\eta_t}}\right)+\sum_{\varsigma=1}^{N_r-1-\lambda}\frac{(\varsigma-1)!}{\left(-{\eta_t}\right)^{N_r-1-\lambda-\varsigma}}\right], i=2,3,\dots,L
\end{split}
\end{equation}
\hrulefill
\end{figure*}
After substituting PDF of $\gamma_i$ given (22) into (39), we formulate the ergodic capacity of the user $i$ with some algebraic manipulations and the aid of \cite [Eq. (4.333.5)]{Gradshteyn1994} as in (40)(see top of the page). Where $\mathbf{Ei}(.)$ and $\Gamma(.)$ are the exponential integral and the gamma function, respectively. $\eta_1=\sum_{p=i}^{L}a_p\rho\sigma_i^2$ and $\eta_2=\sum_{p=i+1}^{L}a_p\rho\sigma_i^2$ are defined. 

Then the ergodic sum rate of SSK-NOMA is obtained as
\begin{equation}
C^{sum}=\sum_{i=1}^{L}C_i
\end{equation}
\subsection{Outage Probability Analysis}
In this section, average outage probabilities for users are presented. Outage event is defined as the situation that the achievable rate of user cannot fulfill the quality of service (QoS) requirement of the user. 
\subsubsection{Outage Probability for first user ($U_1$)}
Outage events of the users are defined as
\begin{equation}
P_r(R_i<\acute{R}_i)
\end{equation}
where $\acute{R}_i$ is the target rate/QoS of the $i$ th user.

For the first user, by substituting (34) into (42)
\begin{equation}
\begin{split}
P_i(out)&=P_r\left(\log_{2}{N_t}\left[1-P_1(e|\mathbf{h_{1,v}})\right]<\acute{R}_1\right) \\
&=P_r\left(P_1(e|\mathbf{h_{1,v}})\geq\psi_1\right)
\end{split}
\end{equation}
where $\psi_1=1-\frac{\acute{R}_1}{\log_{2}{N_t}}$. The average outage probability is given

\begin{equation}
\begin{split}
P_1(out)&=\int\limits_{\psi_1}^{\infty}{P_1(e|\gamma_1)p_{\gamma_1}(\gamma_1)d\gamma_1} \\
&=\int\limits_{0}^{\infty}{P_1(e|\gamma_1)p_{\gamma_1}(\gamma_1)d\gamma_1}-\int\limits_{0}^{\psi_1}{P_1(e|\gamma_1)p_{\gamma_1}(\gamma_1)d\gamma_1}
\end{split}
\end{equation}
The second integral in (44) cannot be solved in a closed-form to the best of our knowledge. Nevertheless, it can be calculated by numerical tools. Furthermore, one can easily see from (44), the outage probability of the $U_1$ is equal to ABEP of $U_1$ (11) when $\acute{R}_1=\log_2{N_t}$ which is expected as result of (35) and (42). 
\subsubsection{Outage Probability for users $i\geq2$}
For users $i\geq2$ by substituting achievable rate of users (36) into (42), we obtain the outage probability 
\begin{equation}
P_i(out)=P_r(SINR_i<\phi_i)
\end{equation}
where $\phi_i=2^{\acute{R}_i-1}$. However, the outage event in (45) considers the perfect SIC for the users. The outage event for the $i$ th user also occurs when the $m$ th user (i.e., $m<i$) signals can not be detected at the $i$ th user. Hence,
\newtheorem{theorem}{Theorem}
\begin{theorem}
We formulate the outage probability of the $i$ th user
 \begin{equation}
\begin{split}
P_i(out)=&P_r(\gamma_i<\psi_i) \\
=&F_{\gamma_i}(\psi_i)
\end{split}
\end{equation}
\end{theorem}
\begin{IEEEproof}
See Appendix D
\end{IEEEproof}
where $\psi_i$ is defined as (47) (see the top of the next page).
\begin{figure*}[!t] 
 \begin{equation}
\psi_i=max\left(\frac{\phi_i}{a_i-\sum_{p=i+1}^{L}a_p\phi_i},\dots \frac{\phi_{m}}{a_{m}-\sum_{p=m+1}^{L}a_p\phi_{m}},\dots \frac{\phi_{2}}{a_{2}-\sum_{p=3}^{L}a_p\phi_{2}}\right),\ m>i,\ i=2,3\dots,L
\end{equation}
\hrulefill
\end{figure*}
Then, we derive the average outage probability of NOMA users (i.e., $i\geq2$) by using the CDF of $\gamma_i$ given in \cite[Eq. (2.3-24)]{Proakis2008} and it turns out to be
\begin{equation}
P_i(out)=1-{e^{\sfrac{-\psi_i}{\overline{\gamma}_i}}\sum_{\lambda=1}^{N_r}\frac{\left(\sfrac{\psi_i}{\overline{\gamma}_i}\right)^{\lambda-1}}{\left(\lambda-1\right)!}}.
\end{equation}
where ${\overline{\gamma}_i}=\rho\sigma_i^2$ is defined.
\begin{figure}[!t]
    \centering
    \includegraphics[width=9cm,height=5cm]{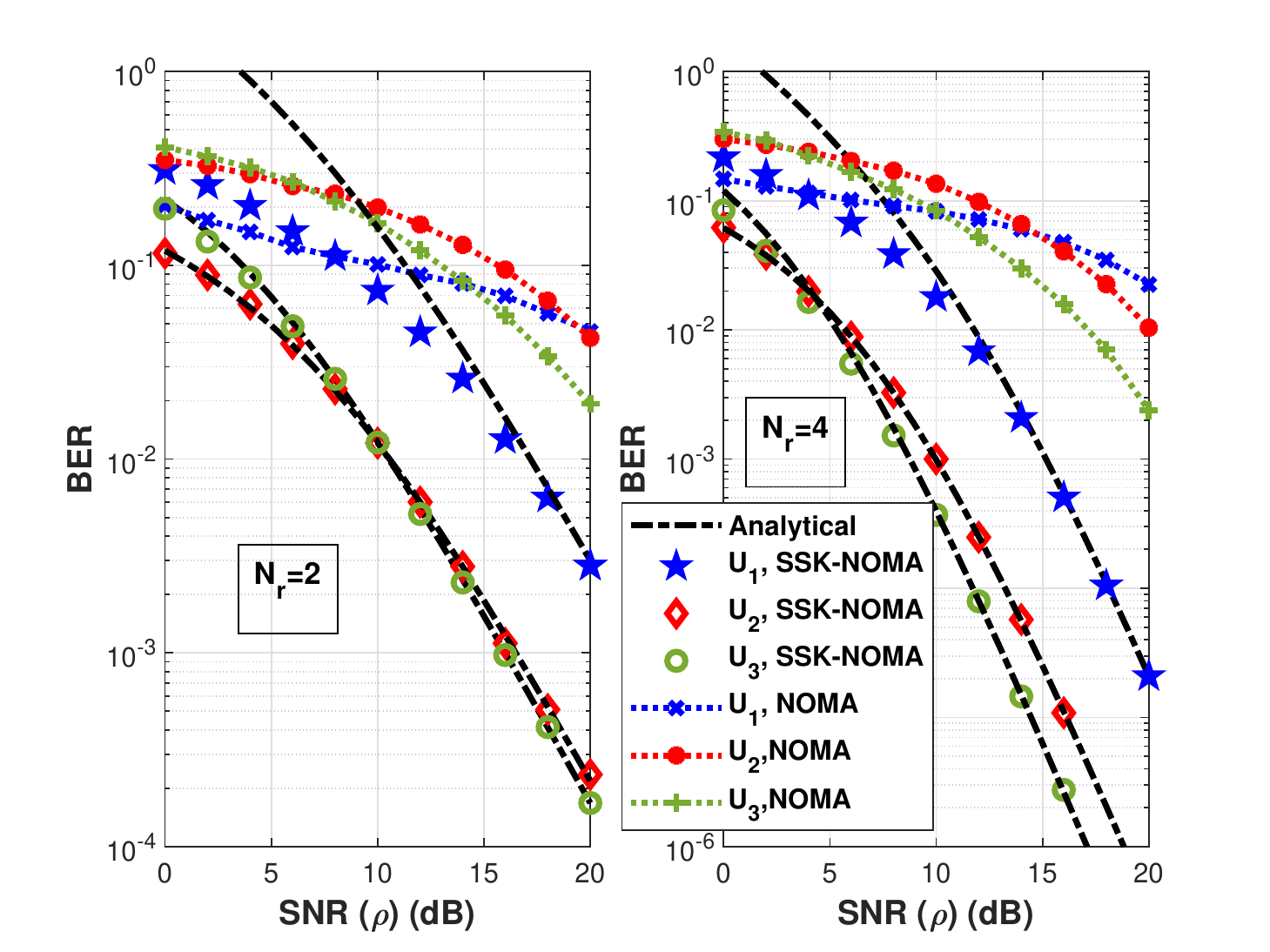}
    \caption{BER comparison of SSK-NOMA and conventional NOMA when $L=3$, and $N_r=2,4$}
    \label{fig2}
\end{figure}
\section{Numerical Results}
In this section we evaluate the derived expressions in Section III via Monte Carlo simulations. Unless otherwise stated, in all figures, lines denote analytical results whereas simulations are denoted by markers and the $N_t=N_r$ is chosen for SSK-NOMA. In all simulations, we assume that $\sigma_i^2=2\sigma_{i-1}^2$ and $\sigma_1^2=0 dB$. The PA coefficients for NOMA users are fixed and chosen as $a=[0.8, \ 0.2]$, $a=[0.7, \ 0.2, \ 0.1]$, $a=[0.6, \ 0.25, \ 0.1, \ 0.05]$ and $a=[0.4, \ 0.25, \ 0.2, \ 0.1, \ 0.05]$ when the number of NOMA users is equal to, 2, 3, 4 and 5, respectively.

In Fig.2, we present the BER performances for users when $L=3$ and  $N_r=2,4$ are set. QPSK is chosen for user $i\geq2$. It is worth pointing out that exact BEP analysis match perfectly with simulations for SSK-NOMA. In addition we provide conventional NOMA simulations to emphasize the superiority of SSK-NOMA. We reveal that SSK-NOMA outperforms NOMA significantly for all users and all users have full diversity order i.e. $N_r$ which can easily seen from the slope of the curves. It is worth pointing out that performance gain for cell-edge user becomes much more than performance gain between conventional SSK and OMA networks provided in \cite{Jeganathan2009}, since with the increase of number of users in NOMA, IUI becomes dominant and this causes poor BER performance. Then, in order to evaluate analytical BER union bound, we provide BER performances of users i.e $i\geq2$ in Fig. 3 when $L=4$ and $N_r=2,4$. One can easily see that the provided upper bound expressions match well with simulations. 

In Fig. 4, ergodic sum rates are presented for SSK-NOMA to uncover the effect of the number of users and the number of antennas. $L=4,5$ and $N_r=2,4,8$ are assumed. In addition, to emphasize superiority of SSK-NOMA, ergodic capacity comparison of SSK-NOMA and conventional NOMA for all users and sum rate is also provided in Fig. 5 when $L=3$ and $N_r=4$. Although, conventional NOMA is proposed as a spectral efficient technique for the next wireless technologies, we reveal that SSK-NOMA offers higher spectral efficiency for all users and overall system since it encounters limited IUI compared to conventional NOMA.  
\begin{figure}[!t]
    \centering
    \includegraphics[width=9cm,height=5cm]{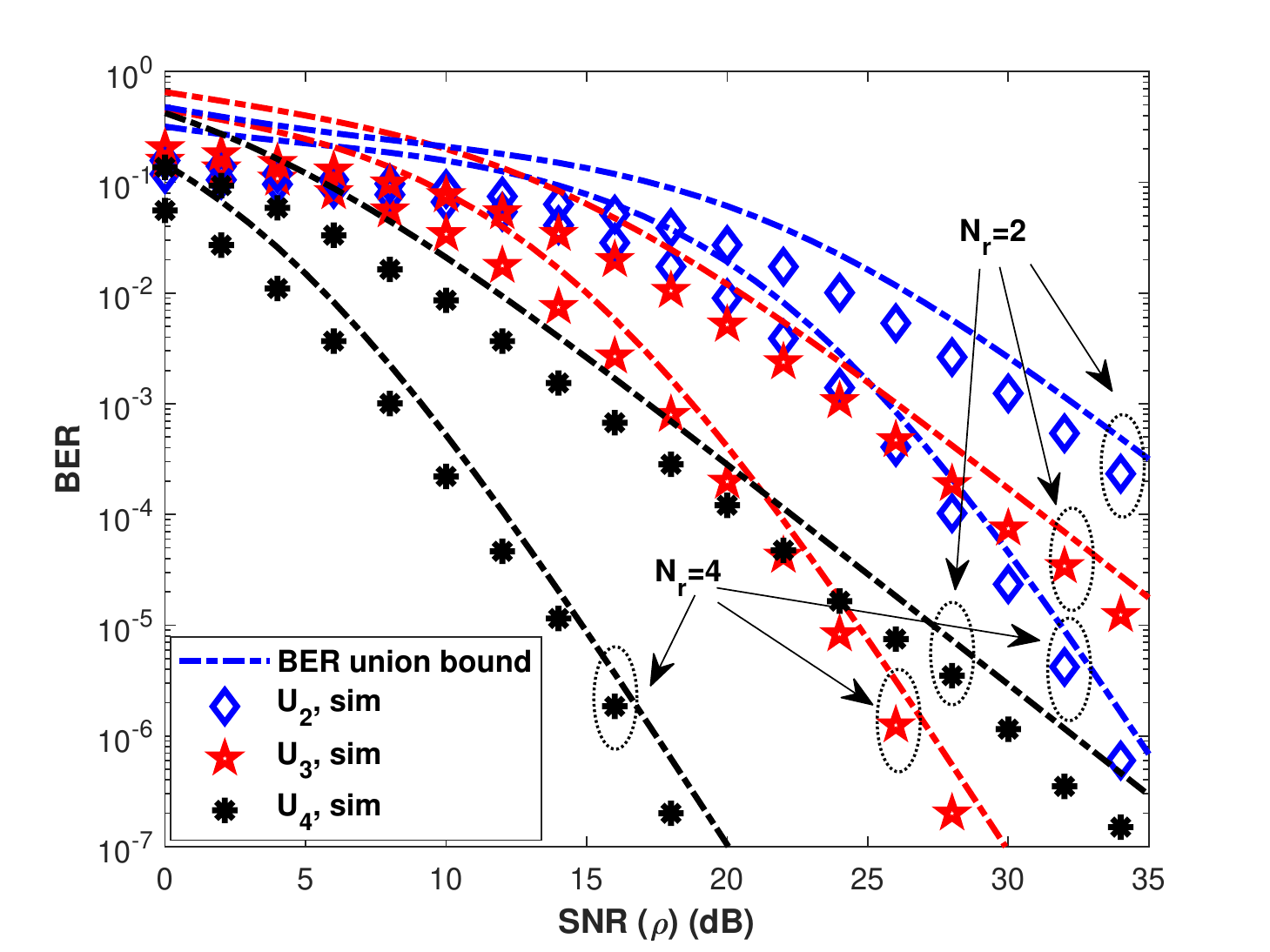}
    \caption{BER union bound of SSK-NOMA for users i.e $i\geq2$ when $L=4$, $N_r=2,4$ }
    \label{fig3}
\end{figure}

\begin{figure}[!t]
    \centering
    \includegraphics[width=9cm,height=5cm]{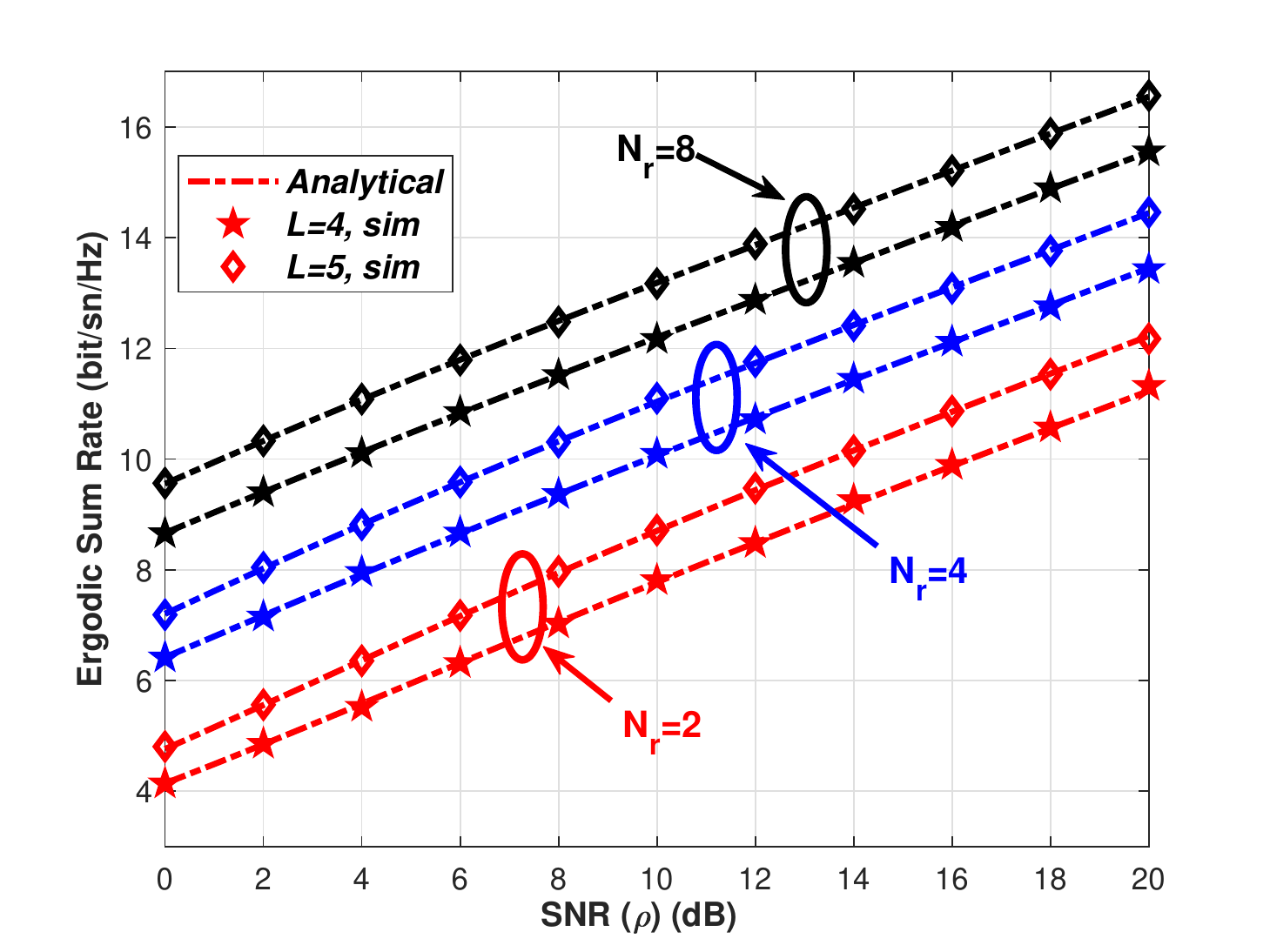}
    \caption{Ergodic sum rate of SSK-NOMA $L=4,5$ and $N_r=2,4,8$ }
    \label{fig4}
\end{figure}

\begin{figure}[!t]
    \centering
    \includegraphics[width=9cm,height=5cm]{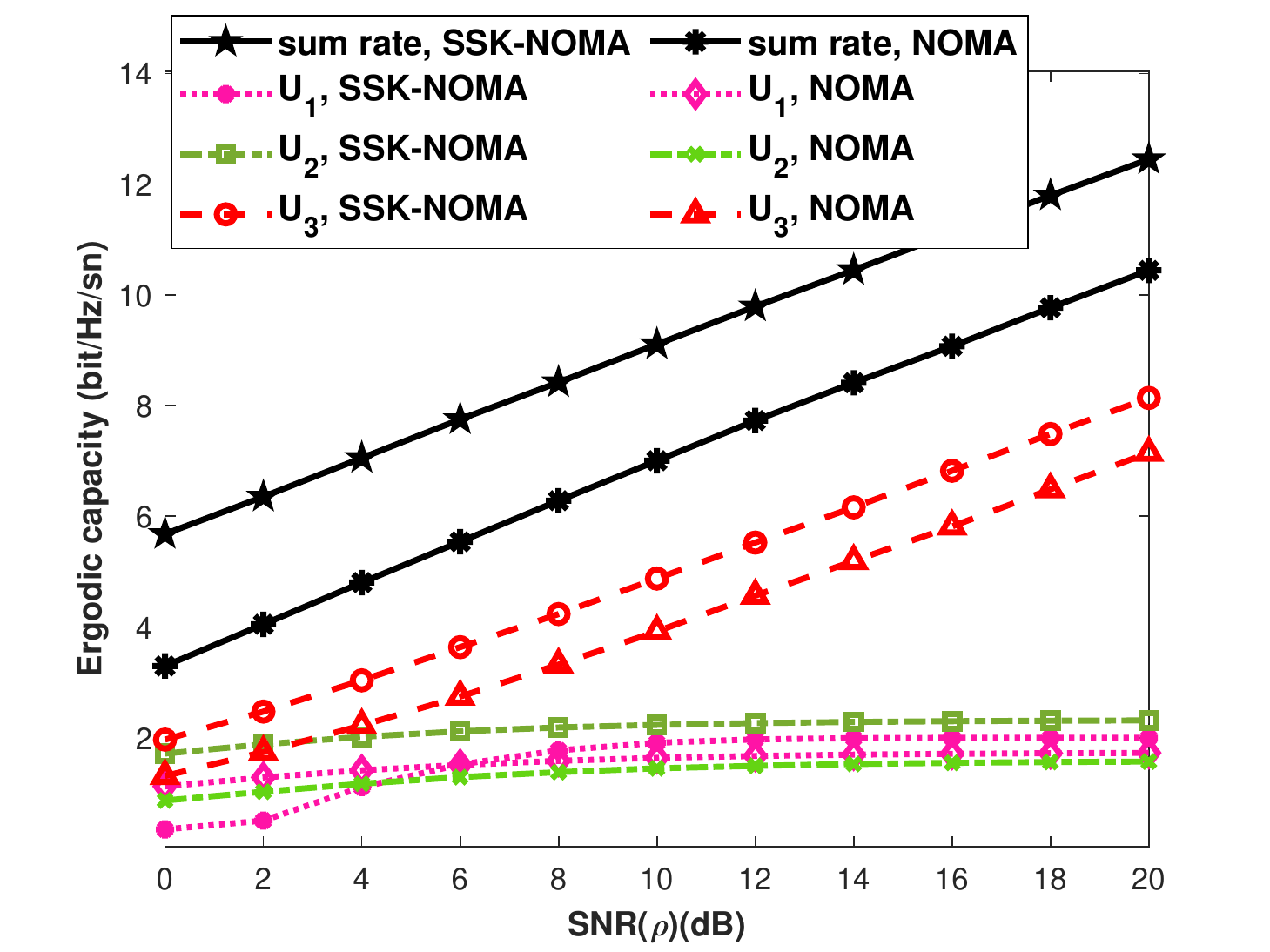}
    \caption{Ergodic capacity comparison of SSK-NOMA and conventional NOMA when $L=3$, $N_r=4$ }
    \label{fig5}
\end{figure}
The outage performances of SSK-NOMA are presented in Fig. 6 for users i.e. $i\geq2$ when $L=4$ and $N_r=2,4$. The target rates of users are chosen, $\acute{R}_2= 1.5 $, $\acute{R}_3=1.5 $ and $\acute{R}_4=2$. The outage performance of $U_1$ is equal to BER performance of user. The analytical derivations match perfectly with simulations. In addition, to compare the outage performances of SSK-NOMA and conventional NOMA, simulations are presented in Fig. 7 for two different target rates when $L=3$ and $N_r=2$. In the first case, the target rates $\acute{R}_1=1$, $\acute{R}_2= 1 $, $\acute{R}_3=2 $ are chosen and SSK-NOMA outperforms conventional NOMA  for $U_2$ and $U_3$. Although NOMA seems to be very few superior to SSK-NOMA for $U_1$ in this case, in the second case, the target rates are increased to $\acute{R}_1=2$, $\acute{R}_2= 2$, $\acute{R}_3=2.5$, and conventional NOMA remains in outage for all users until $10 dB$ whereas the SSK-NOMA has similar performance to previous case. We note that the target rate of $U_1$ in SSK-NOMA is upgraded by just implementing more transmit antennas without any increase in complexity cost.

\begin{figure}[!t]
    \centering
    \includegraphics[width=9cm,height=5cm]{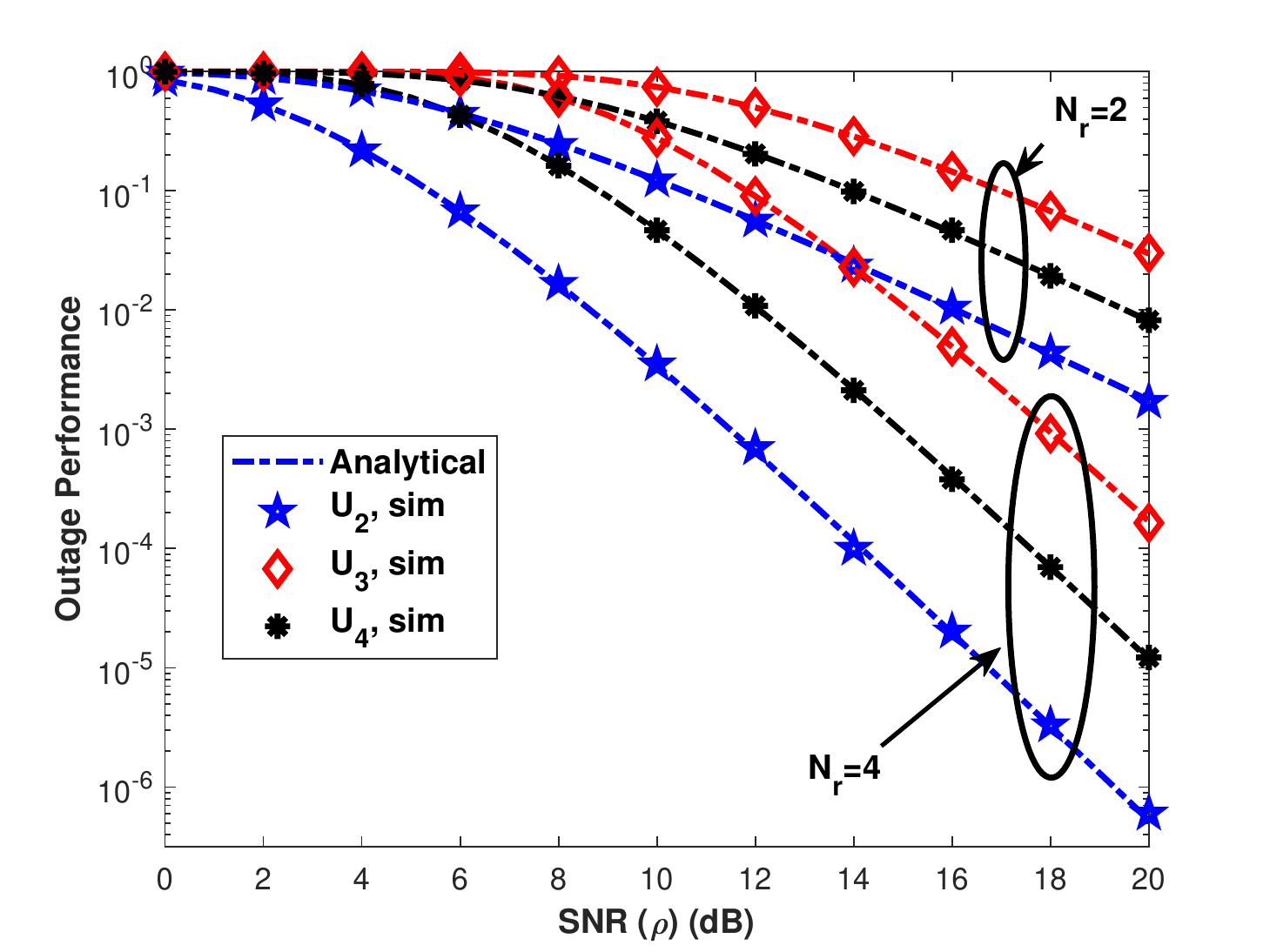}
    \caption{Outage performance of SSK-NOMA when $L=4$, $N_r=2,4$, $\acute{R}_2= 1.5 $, $\acute{R}_3=1.5 $ and $\acute{R}_4=2$ }
    \label{fig6}
\end{figure}

\begin{figure}[!t]
    \centering
    \includegraphics[width=9cm,height=5cm]{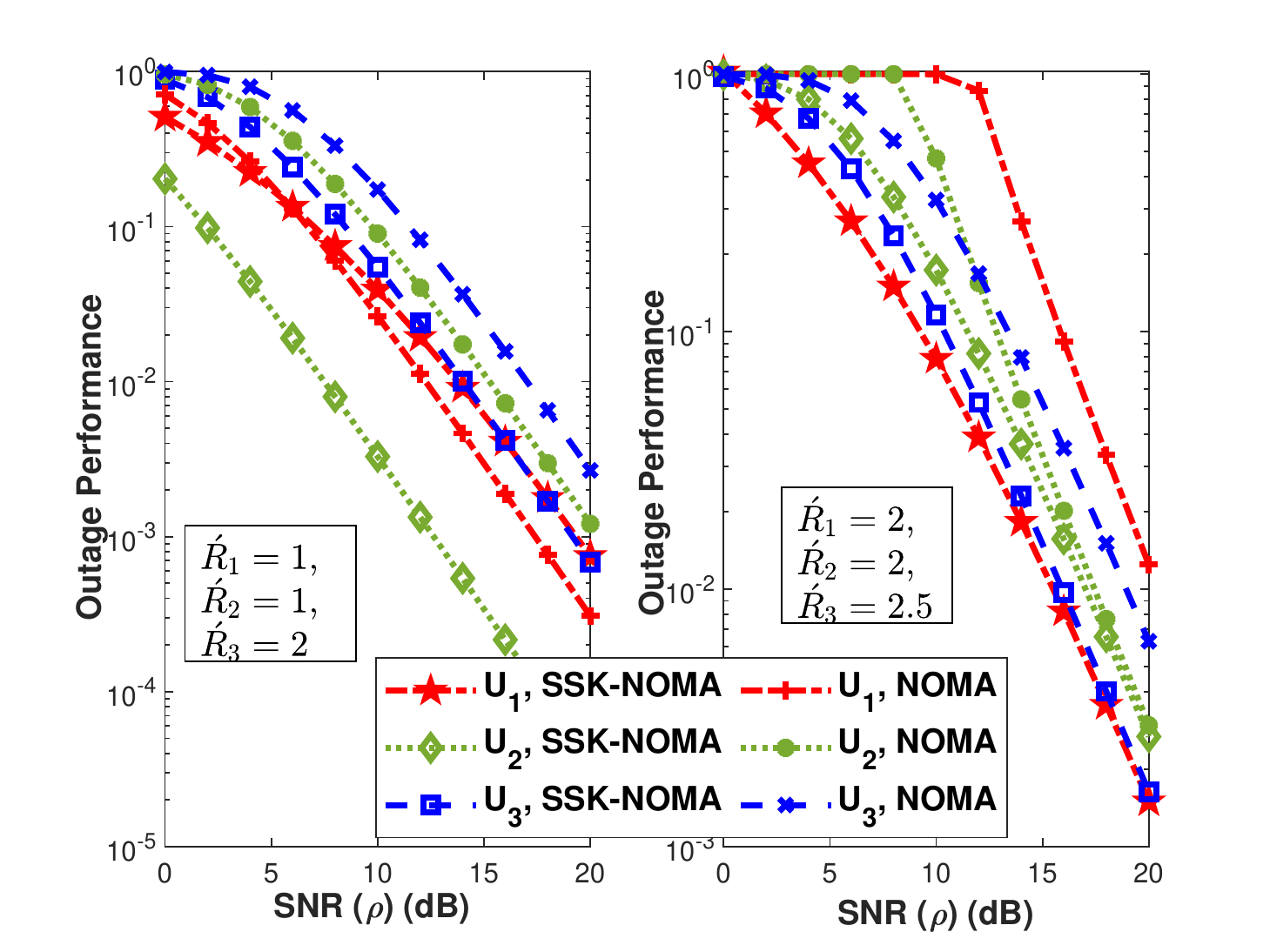}
    \caption{Outage comparison of SSK-NOMA and conventional NOMA when $L=3$, $N_r=2$ }
    \label{fig7}
\end{figure}
Finally, we investigate the effect of PA coefficients on the performance of SSK-NOMA. To provide that, we present the BER and outage performance of users with the change of $a_2$ for $L=3$ at $\rho=20dB$ SNR in Fig. 8 and Fig. 9, respectively. The BER and outage performance of $U_1$ become the same when the $\acute{R}_1=\log_2{N_t}$, hence the the outage performance of $U_1$ is not provided in Fig. 9. The performances of users in Fig. 8 and Fig. 9 reveal that the wrong chosen of PA coefficient causes for NOMA users to communicate with low reliability (poor BER) or to be in outage. Hence, the optimal PA should be adopted according to target rate and target reliability of the system. 
\begin{figure}[!t]
    \centering
    \includegraphics[width=9cm,height=5cm]{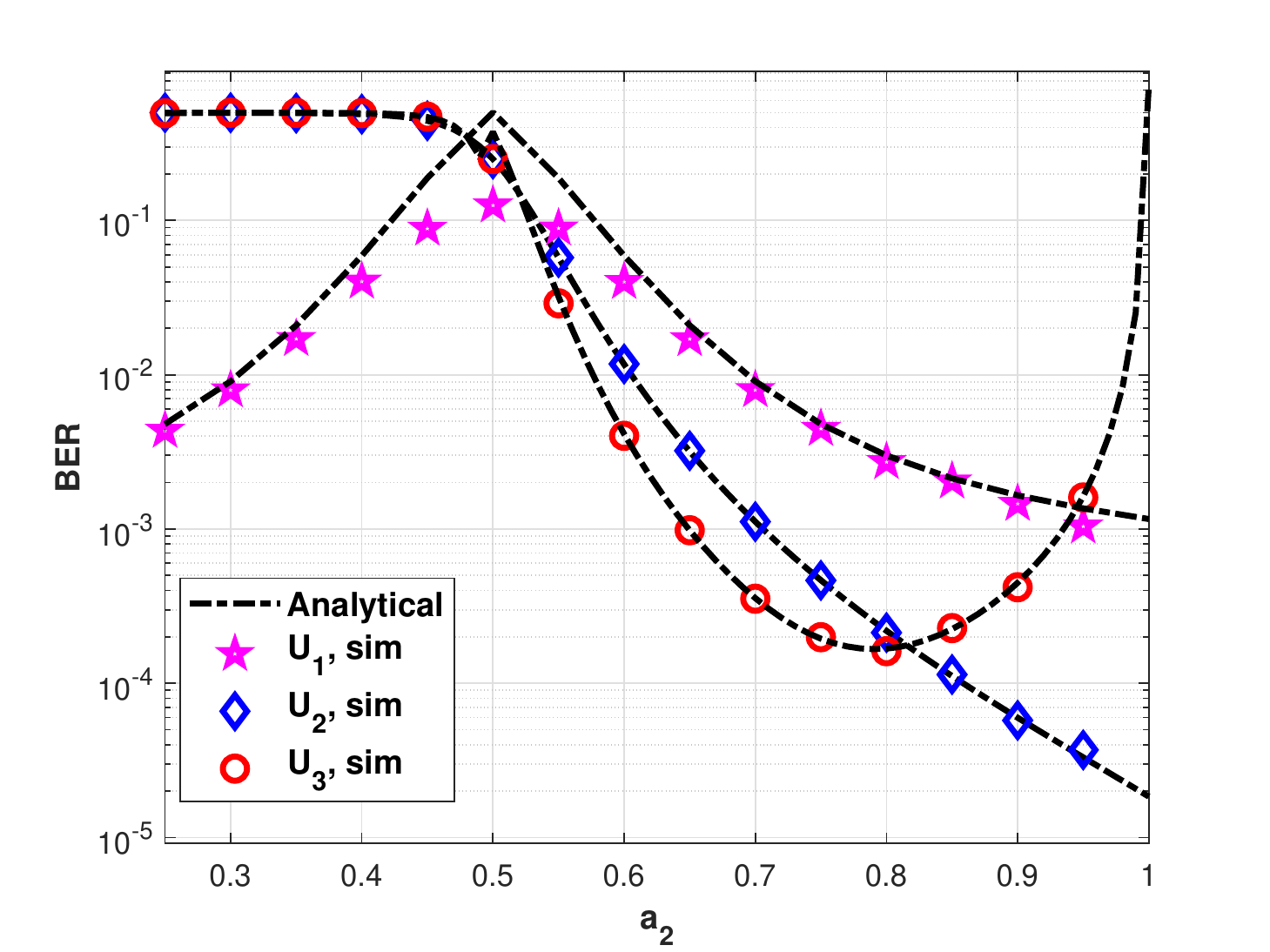}
    \caption{Effect of PA on error performance of SSK-NOMA, $L=3$, $N_r=2$  }
    \label{fig8}
\end{figure}

\begin{figure}[!t]
    \centering
    \includegraphics[width=9cm,height=5cm]{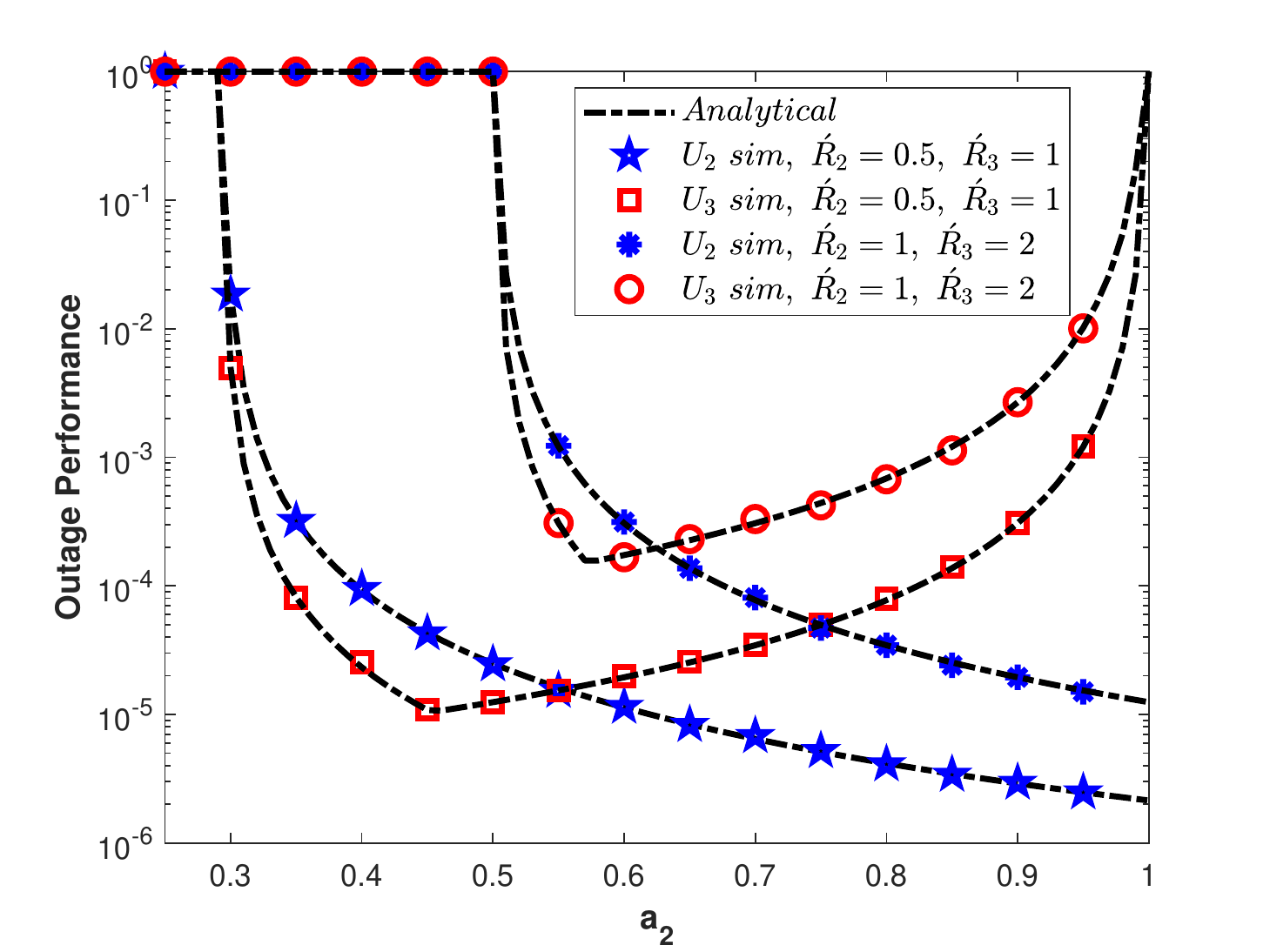}
    \caption{Effect of PA on outage performance of SSK-NOMA, $L=3$, $N_r=2$ }
    \label{fig9}
\end{figure}
\section{Conclusion}
In this paper, we provide a comprehensive analytical framework for SSK-NOMA systems. We derive closed-form expressions for ABEP, ergodic sum rate and outage probability of SSK-NOMA. All derived expressions are validated via simulations. We reveal that, SSK-NOMA outperforms conventional NOMA systems in terms of all performance metrics (i.e., BER, sum rate, outage). In this paper, we assume the fixed-PA usage and then we present the effect of the PA on the BER and outage performance of SSK-NOMA. The results show that the optimal PA for SSK-NOMA can be obtained by considering outage and BER constraints for the chosen rate and reliability target. Since the SSK-NOMA provide better performance and lower energy consumption, the combinations of SSK-NOMA with the other physical techniques such as cooperative communication, energy harvesting and cognitive communication are seen as the future directions of research. The main costs in SSK-NOMA can be listed in two items. Firstly, the receiver complexity at the cell-edge user increases exponentially according to number of users in a resource block. Secondly, the number of required transmit antennas increases with the power of 2 according to cell-edge user's target rate. Recalling that most of NOMA related models limit the number of users in a resource block with only two and resource allocation/user clustering algorithms are investigated. By using these resource allocation algorithms, number of users in a resource block can be limited and then, the complexity of SSK-NOMA will be much less than conventional NOMA networks with the same number of users in a resource block. Nevertheless, even for higher users in a resource block, the complexity of cell-edge user/first-cost is affordable considering the performance gain. Finally, the required transmit antenna number/second-cost can be reduced by implementing other index modulation techniques such as channel/media-based modulation (MBM). Resource allocation for SSK-NOMA and MBM-NOMA are quite promising subjects for future researchers.
\appendices
%\section{Proof of the exact BEP analysis for $L=3$}

\section{}%{Proof of BEP of $U_2$}

Considered the QPSK is used for both users with different PA coefficients, the base-band signal space of the SC symbols at BS is given Table II. In Table II, the binary symbols of the users are given in the form $b_{i,l}b_{i,l}$  where the first index denotes the user $i$ and the second index represents the $l$ th bit of the $i$ th user. 
\begin{table*}[!t]
  \centering
  \caption{\textsc{Base-band SC symbols for L=3 and QPSK at BS}}
  \label{table2}
  \begin{tabular}{|c|c|c|c|c|c|}
	\hline
	&&{$b_{3,2}b_{3,1}=00$}&{$b_{3,2}b_{3,1}=01$}&{$b_{3,2}b_{3,1}=10$}&{$b_{3,2}b_{3,1}=11$} \\
	\hline
	\multirow{2}{*}{$b_{2,2}b_{2,1}=00$}&Re&$\sqrt{\sfrac{a_2}{2}}+\sqrt{\sfrac{a_3}{2}}$&$\sqrt{\sfrac{a_2}{2}}-\sqrt{\sfrac{a_3}{2}}$&$\sqrt{\sfrac{a_2}{2}}+\sqrt{\sfrac{a_3}{2}}$&$\sqrt{\sfrac{a_2}{2}}-\sqrt{\sfrac{a_3}{2}}$ \\ \cline{2-6}
	&Im&$\sqrt{\sfrac{a_2}{2}}+\sqrt{\sfrac{a_3}{2}}$&$\sqrt{\sfrac{a_2}{2}}+\sqrt{\sfrac{a_3}{2}}$&$\sqrt{\sfrac{a_2}{2}}-\sqrt{\sfrac{a_3}{2}}$&$\sqrt{\sfrac{a_2}{2}}-\sqrt{\sfrac{a_3}{2}}$ \\
	\hline
\multirow{2}{*}{$b_{2,2}b_{2,1}=01$}&Re&$-\sqrt{\sfrac{a_2}{2}}+\sqrt{\sfrac{a_3}{2}}$&$-\sqrt{\sfrac{a_2}{2}}-\sqrt{\sfrac{a_3}{2}}$&$-\sqrt{\sfrac{a_2}{2}}+\sqrt{\sfrac{a_3}{2}}$&$-\sqrt{\sfrac{a_2}{2}}-\sqrt{\sfrac{a_3}{2}}$ \\ \cline{2-6}
	&Im&$\sqrt{\sfrac{a_2}{2}}+\sqrt{\sfrac{a_3}{2}}$&$\sqrt{\sfrac{a_2}{2}}+\sqrt{\sfrac{a_3}{2}}$&$\sqrt{\sfrac{a_2}{2}}-\sqrt{\sfrac{a_3}{2}}$&$\sqrt{\sfrac{a_2}{2}}-\sqrt{\sfrac{a_3}{2}}$ \\
	\hline
	\multirow{2}{*}{$b_{2,2}b_{2,1}=10$}&Re&$\sqrt{\sfrac{a_2}{2}}+\sqrt{\sfrac{a_3}{2}}$&$\sqrt{\sfrac{a_2}{2}}-\sqrt{\sfrac{a_3}{2}}$&$\sqrt{\sfrac{a_2}{2}}+\sqrt{\sfrac{a_3}{2}}$&$\sqrt{\sfrac{a_2}{2}}-\sqrt{\sfrac{a_3}{2}}$ \\ \cline{2-6}
	&Im&$-\sqrt{\sfrac{a_2}{2}}+\sqrt{\sfrac{a_3}{2}}$&$-\sqrt{\sfrac{a_2}{2}}+\sqrt{\sfrac{a_3}{2}}$&$-\sqrt{\sfrac{a_2}{2}}-\sqrt{\sfrac{a_3}{2}}$&$-\sqrt{\sfrac{a_2}{2}}-\sqrt{\sfrac{a_3}{2}}$ \\
	\hline
\multirow{2}{*}{$b_{2,2}b_{2,1}=11$}&Re&$-\sqrt{\sfrac{a_2}{2}}+\sqrt{\sfrac{a_3}{2}}$&$-\sqrt{\sfrac{a_2}{2}}-\sqrt{\sfrac{a_3}{2}}$&$-\sqrt{\sfrac{a_2}{2}}+\sqrt{\sfrac{a_3}{2}}$&$-\sqrt{\sfrac{a_2}{2}}-\sqrt{\sfrac{a_3}{2}}$ \\ \cline{2-6}
	&Im&$-\sqrt{\sfrac{a_2}{2}}+\sqrt{\sfrac{a_3}{2}}$&$-\sqrt{\sfrac{a_2}{2}}+\sqrt{\sfrac{a_3}{2}}$&$-\sqrt{\sfrac{a_2}{2}}-\sqrt{\sfrac{a_3}{2}}$&$-\sqrt{\sfrac{a_2}{2}}-\sqrt{\sfrac{a_3}{2}}$ \\
	\hline
  \end{tabular}
\end{table*}
One can easily see the transmitted symbols have different energy levels with the different probabilities. Since the $U_2$ implements only ML detection by treating $U_3$ symbols as noise, the ML decision boundary for each dimension is given as $\mathbf{r_2^I}<0$ or $\mathbf{r_2^I}\geq0$ and $\mathbf{r_2^R}<0$ or $\mathbf{r_2^R}\geq0$. The superscripts $()^I$ and $()^R$ denote imaginary and real components of the base-band signal.

According to ML decision rule, the error probability for the $first$ bit of the $U_2$ is obtained as
\begin {equation}
\begin{split}
P_{2,1}(e|\mathbf{h_{2,v}})= &\frac{1}{2}P_r(\mathbf{w_2^R}\geq{\left(\sqrt{\sfrac{a_2}{2}}+\sqrt{\sfrac{a_3}{2}}\right)}\sqrt{P}\mathbf{h_{2,v}}\mathbf{h_{2,v}^H})\\
&+\frac{1}{2}P_r(\mathbf{w_2^R}\geq{\left(\sqrt{\sfrac{a_2}{2}}-\sqrt{\sfrac{a_3}{2}}\right)}\sqrt{P}\mathbf{h_{2,v}}\mathbf{h_{2,v}^H})
\end{split}
\end{equation}
where $\mathbf{w_2^R}$ is the real part of the AWGN and it has zero mean with $\sfrac{N_0}{2}$ variance. Hence, the conditional BEP for the $first$ bit of the $U_2$ is obtained
\begin{equation}
\begin{split}
\ P_{2,1}(e|\gamma_2)=&\frac{1}{2}\left[Q\left( \sqrt{\left(\sqrt{{a_2}}+\sqrt{{a_3}}\right)^2\gamma_2}\right) \right. \\
&\left.+Q\left( \sqrt{\left(\sqrt{{a_2}}-\sqrt{{a_3}}\right)^2\gamma_2}\right)\right] 
\end{split}
\end{equation}
The conditional BEP for the $second$ bit of $U_2$ can be easily obtained by considering ML decision rule and $\mathbf{w_2^I}$. One can easily see that the BEP for the $second$ bit of $U_2$ is the same with (50). Hence the BEP of $U_2$ is obtained as in (21) by formulating $P_2(e)=\sfrac{\left(P_{2,1}(e)+P_{2,2}(e)\right)}{2}$. Proof is completed.

\section{}%{Proof of BEP of $U_3$ under condition $U_2$ detected correctly}
Under the condition that the $U_2$ symbols are detected correctly and subtracted from the received signal, a regular QPSK constellation with the symbol energy $a_3\rho$ is remained. Nevertheless, we emphasize that this  is a conditional case which depends on the correct detection of $U_2$ symbols. We easily obtain the correct detection probability of $U_2$ symbols at $U_3$ by using (49) when the channel gains are changed with $\mathbf{h_{3,v}}$. Without loss off generality, we assume that $b_{2,2}b_{2,1}=00$ is sent and detected correctly. In this case, the conditional probability for the $first$ bit of $U_3$ on $\mathbf{w_3}$ including priori probability of correct detection of $U_2$ is given (51) (see the top of the next page)
\begin{figure*}
\begin{equation}
\begin{split}
\ &P_{3,1}(e|_{correct_{U_2},\mathbf{w_2^R}})= \frac{1}{2}P_r(\mathbf{w_3^R}\geq{-\left(\sqrt{\sfrac{a_2}{2}}+\sqrt{\sfrac{a_3}{2}}\right)}\sqrt{P}\mathbf{h_{3,v}}\mathbf{h_{3,v}^H}) \times \\
&P_r(\mathbf{w_3^R}<{-\left(\sqrt{\sfrac{a_3}{2}}\right)}\sqrt{P}\mathbf{h_{2,v}}\mathbf{h_{2,v}^H}|\mathbf{w_3^R}\geq{-\left(\sqrt{\sfrac{a_2}{2}}+\sqrt{\sfrac{a_3}{2}}\right)}\sqrt{P}\mathbf{h_{3,v}}\mathbf{h_{3,v}^H}+\frac{1}{2}P_r(\mathbf{w_3^R}\geq-{\left(\sqrt{\sfrac{a_2}{2}}-\sqrt{\sfrac{a_3}{2}}\right)}\sqrt{P}\mathbf{h_{3,v}}\mathbf{h_{3,v}^H})\\
&\times P_r(\mathbf{w_3^R}\geq{\left(\sqrt{\sfrac{a_3}{2}}\right)}\sqrt{P}\mathbf{h_{3,v}}\mathbf{h_{3,v}^H}|\mathbf{w_3^R}\geq-{\left(\sqrt{\sfrac{a_2}{2}}-\sqrt{\sfrac{a_3}{2}}\right)}\sqrt{P}\mathbf{h_{3,v}}\mathbf{h_{3,v}^H})
\end{split}
\end{equation}
\hrulefill
\end{figure*}

Recalling $\gamma_i=\rho\mathbf{h_{i,v}}\mathbf{h_{i,v}^H}$ and by utilizing the conditional probability $P_r(A|B)=\frac{P_r(A\cap B)}{P_r(B)}$ and after some algebraic manipulations, we obtain the BEP for the $first$ bit of $U_3$ under the condition that $U_2$ symbols detected correctly
\begin{equation}
\begin{split}
P_{3,1}(e|correct_{U_2})=&\frac{1}{2}\left[2Q\left(\sqrt{a_3\gamma_3}\right)\right.\\
&\left.-Q\left(\sqrt{\left(\sqrt{a_2}+\sqrt{a_3}\right)^2\gamma_3}\right)\right]
\end{split}
\end{equation}
The BEP for the $second$ bit of $U_3$ is obtained as same as (52) by replacing the AWGN $\mathbf{w_2}^I$ and the total BEP of $U_3$ is obtained as in (25). The proof is completed.

\section{}%{Proof of BEP of $U_3$ under condition $U_2$ detected erroneously}

In order to obtain BEP of the second case, without loss of generality we assume that binary symbol of $U_2$ $b_{2,2}b_{2,1}=00$ is transmitted and detected as $b_{2,2}b_{2,1}=11$ at user $U_3$ and subtracted from the received signal. The obtained signal after SIC is given by $\mathbf{r'_3}=\Upsilon\sqrt{P}\mathbf{h_{3,v}}+\mathbf{w_{3}}$. The base-band signal constellation $\Upsilon$ is given in Table III. 

\begin{table}[!ht]
  \centering
  \caption{\textsc{Base-band symbols at $U_3$ after erroneous detection of $U_2$}}
  \label{table3}
  \begin{tabular}{|c|c|c|}
	\hline
	\multirow{2}{*}{}&\multicolumn{2}{c|}{$\Upsilon$}\\ \cline{2-3}
	{}&Re&Im \\
	\hline
	{$b_{3,2}b_{3,1}=00$}&$2\sqrt{\sfrac{a_2}{2}}+\sqrt{\sfrac{a_3}{2}}$&$2\sqrt{\sfrac{a_2}{2}}+\sqrt{\sfrac{a_3}{2}}$\\
	\hline
		{$b_{3,2}b_{3,1}=01$}&$2\sqrt{\sfrac{a_2}{2}}-\sqrt{\sfrac{a_3}{2}}$&$2\sqrt{\sfrac{a_2}{2}}+\sqrt{\sfrac{a_3}{2}}$\\
	\hline
		{$b_{3,2}b_{3,1}=10$}&$2\sqrt{\sfrac{a_2}{2}}+\sqrt{\sfrac{a_3}{2}}$&$2\sqrt{\sfrac{a_2}{2}}-\sqrt{\sfrac{a_3}{2}}$\\
	\hline
	{$b_{3,2}b_{3,1}=11$}&$2\sqrt{\sfrac{a_2}{2}}-\sqrt{\sfrac{a_3}{2}}$&$2\sqrt{\sfrac{a_2}{2}}-\sqrt{\sfrac{a_3}{2}}$\\
	\hline
\end{tabular}
\end{table}
As in the previous case, the BEP for the $first$ bit of $U_3$ is obtained by considering the condition on $\mathbf{w_3}^R$. Including priori probability of error detection of $U_2$, th BEP for the $first$ bit of $U_3$ is given in (53) (see the top of the next page).
\begin{figure*}[!t] 
\begin{equation}
\begin{split}
\ P_{3,1}(e|_{error_{U_2},\mathbf{w_3^R}} )= &\frac{1}{2}P_r(\mathbf{w_3^R}<{-\left(\sqrt{\sfrac{a_2}{2}}+\sqrt{\sfrac{a_3}{2}}\right)}\sqrt{P}\mathbf{h_{3,v}}\mathbf{h_{3,v}^H})\\
&\times P_r(\mathbf{w_3^R}<{-\left(2\sqrt{\sfrac{a_2}{2}}+\sqrt{\sfrac{a_3}{2}}\right)}\sqrt{P}\mathbf{h_{2,v}}\mathbf{h_{2,v}^H}|\mathbf{w_3^R}<{-\left(\sqrt{\sfrac{a_2}{2}}+\sqrt{\sfrac{a_3}{2}}\right)}\sqrt{P}\mathbf{h_{3,v}}\mathbf{h_{3,v}^H})\\
&+\frac{1}{2}P_r(\mathbf{w_3^R}<-{\left(\sqrt{\sfrac{a_2}{2}}-\sqrt{\sfrac{a_3}{2}}\right)}\sqrt{P}\mathbf{h_{3,v}}\mathbf{h_{3,v}^H})\\
&\times P_r(\mathbf{w_3^R}\geq{-\left(2\sqrt{\sfrac{a_2}{2}}-\sqrt{\sfrac{a_3}{2}}\right)}\sqrt{P}\mathbf{h_{3,v}}\mathbf{h_{3,v}^H}|\mathbf{w_3^R}<-{\left(\sqrt{\sfrac{a_2}{2}}-\sqrt{\sfrac{a_3}{2}}\right)}\sqrt{P}\mathbf{h_{3,v}}\mathbf{h_{3,v}^H})
\end{split}
\end{equation}
\hrulefill
\end{figure*}
Again by utilizing $P_r(A|B)=\frac{P_r(A\cap B)}{P_r(B)}$ and after some algebraic manipulations, we obtain the BEP for the $first$ bit of $U_3$ under the condition that $U_2$ symbols detected erroneously
\begin{equation}
\begin{split}
&P_{3,1}(e|error_{U_2})=\frac{1}{2}\left[Q\left(\sqrt{\left(\sqrt{a_2}-\sqrt{a_3}\right)^2\gamma_3}\right)\right.\\
&\left.-Q\left(\sqrt{\left(2\sqrt{a_2}-\sqrt{a_3}\right)^2\gamma_3}\right)+Q\left(\sqrt{\left(2\sqrt{a_2}+\sqrt{a_3}\right)^2\gamma_3}\right)\right]
\end{split}
\end{equation}
It is seen that the BEP for the $second$ bit is obtained as same as (54) when AWGN is changed with $\mathbf{w_3^I}$. By averaging the BEP for each bit of $U_3$, the expression given (26) is obtained so the proof is completed. 
\section{}
Considered that the imperfect SIC, the outage event of the $i$ th user turns out to be as (55)(see the top of the next page)

\begin{figure*}[!t] 
\begin{equation}
P_i(out)=P_r(SINR_i<\phi_i)\cup P_r(SINR_{i\rightarrow i-1}<\phi_{i-1})\cup \dots \cup P_r(SINR_{i\rightarrow m}<\phi_{m})\cup\dots\cup P_r(SINR_{i\rightarrow 2}<\phi_2)
\end{equation}
\hrulefill
\end{figure*}
where $SINR_{i\rightarrow m}$ denotes the SINR for the detecting of $m$ th user at the user $i$ and is defined
 \begin{equation}
SINR_{i\rightarrow m}=\frac{a_m\rho\mathbf{h_{i,v}}\mathbf{h_{i,v}^H}}{\sum_{p=m+1}^{L}a_p\rho\mathbf{h_{i,v}}\mathbf{h_{i,v}^H}+1}
\end{equation}

The outage event given in (55) can be formulated as 

\begin{equation}
F_{i\rightarrow i}(\phi_i)\cup\dots F_{i\rightarrow m}(\phi_i)\cup\dots F_{i\rightarrow 2}(\phi_2),m<i
\end{equation}
where $F_{i\rightarrow m}$ is the cumulative distribution function (CDF) of $SINR_{i\rightarrow m}$. Recalling that $\rho\mathbf{h_{i,v}}\mathbf{h_{i,v}^H}$ is the SNR at the output of MRC at user $i$ 
\begin{equation}
\begin{split}
F_{i\rightarrow m}(\phi_i)&=P_r(\frac{a_m\gamma_i}{1\sum_{p=m+1}^{L}a_p\gamma_i}<\phi_m)\\
&=P_r(\gamma_i<\frac{\phi_{m}}{a_{m}-\sum_{p=m+1}^{L}a_p\phi_{m}})
\end{split}
\end{equation}
and
\begin{equation}
\begin{split}
&P_{i}(out)=F_{\gamma_i}(\frac{\phi_{i}}{a_{i}-\sum_{p=i+1}^{L}a_p\phi_{i}})\cup\dots \\
&F_{\gamma_i}(\frac{\phi_{m}}{a_{m}-\sum_{p=m+1}^{L}a_p\phi_{m}})\dots \cup F_{\gamma_i}(\frac{\phi_{2}}{a_{2}-\sum_{p=3}^{L}a_p\phi_{2}})
\end{split}
\end{equation}
which is expressed in (46) so the proof is completed.

\ifCLASSOPTIONcaptionsoff
  \newpage
\fi



%\bibliographystyle{IEEEtran}
\bibliographystyle{IEEEtran}
\bibliography{ssk_noma_tvt}

\begin{IEEEbiography}{Ferdi Kara}
received B.Sc. (with Hons.) in electronics and communication engineering from Suleyman Demirel University, Turkey in 2011 and M.Sc. in electrical and electronics engineering from Zonguldak Bulent Ecevit University, Turkey, in 2015. He is currently working towards his Ph.D degree. Mr. Kara serves as regular reviewer for several IEEE journals. His main research areas are NOMA, MIMO sytems, cooperative communication and machine learning in physical communication.
\end{IEEEbiography}

\begin{IEEEbiography}{Hakan Kaya}
received B.Sc., M.Sc., and Ph.D degrees in all electrical and electronics engineering from Zonguldak Karaelmas University, Turkey, in 2007, 2010, and 2015, respectively. Since 2015, he has been working at Zonguldak Bulent Ecevit University as Assistant Professor. His research interests are cooperative communication, NOMA, turbo coding and machine learning.
\end{IEEEbiography}

%\newpage


%\vfill


\end{document}